\documentclass[preprint,12pt]{elsarticle}

\usepackage{amssymb}
\usepackage{amsmath}
\usepackage{array}
\usepackage{cancel}
\usepackage{mathtools}
\usepackage{listings}
\usepackage{placeins}
\usepackage{graphicx}
\usepackage{subcaption}

\usepackage{booktabs}
\newcolumntype{C}[1]{>{\centering\let\newline\\\arraybackslash\hspace{0pt}}m{#1}}
\usepackage{colortbl}
\definecolor{C1} {RGB}{224, 224, 224}
\usepackage{mathtools}
\usepackage{siunitx}
\newcolumntype{T}[1]{S[table-format=#1]}
\usepackage{xparse}
\NewDocumentCommand\uvec{}{\mathbf{u}}
\NewDocumentCommand\pvec{}{\mathbf{p}}
\NewDocumentCommand\ptil{}{\Tilde{\pvec}}
\NewDocumentCommand\dt{}{\partial_t}
\NewDocumentCommand\br{m}{\left(#1\right)}
\NewDocumentCommand\sbr{m}{\left[#1\right]}
\NewDocumentCommand\C{O{}O{}O{}}{C_{#1}^{#2}(\uvec_s^{#3})}
\DeclareMathSymbol{\shortminus}{\mathbin}{AMSa}{"39}

\NewDocumentCommand\lr{m}{\langle#1\rangle}

\journal{Journal of Computational Physics}

\begin{document}


\begin{frontmatter}

\title{Quantifying the checkerboard problem to reduce numerical dissipation}
\author[1]{J.A. Hopman\corref{cor1}}
\ead{jannes.hopman@upc.edu}
\ead[url]{github.com/janneshopman}
\author[1]{D. Santos}
\author[1,2]{À. Alsalti-Baldellou}
\author[1]{J. Rigola}
\author[1]{F.X. Trias}
\cortext[cor1]{Corresponding author}
\affiliation[1]{organization={Heat and Mass Transfer Technological Center, Technical University of Catalonia},
            addressline={ESEIAAT, c/Colom 11},
            city={Terrassa},
            postcode={08222},
            state={Barcelona},
            country={Spain}}
\affiliation[2]{organization={Termo Fluids SL},
            addressline={www.termofluids.com},
            city={Sabadell},
            state={Barcelona},
            country={Spain}}

\begin{abstract}
This work provides a comprehensive exploration of various methods in solving incompressible flows using a projection method, and their relation to the occurrence and management of checkerboard oscillations. It employs an algebraic symmetry-preserving framework, clarifying the derivation and implementation of discrete operators while also addressing the associated numerical errors. The lack of a proper definition for the checkerboard problem is addressed by proposing a physics-based coefficient. This coefficient, rooted in the disparity between the compact- and wide-stencil Laplacian operators, is able to quantify oscillatory solution fields with a physics-based, global, normalised, non-dimensional value. The influence of mesh and time-step refinement on the occurrence of checkerboarding is highlighted. Therefore, single measurements using this coefficient should be considered with caution, as the value presents little use without any context and can either suggest mesh refinement or use of a different solver. 

In addition, an example is given on how to employ this coefficient, by establishing a negative feedback between the level of checkerboarding and the inclusion of a pressure predictor, to dynamically balance the checkerboarding and numerical dissipation. This method is tested for laminar and turbulent flows, demonstrating its capabilities in obtaining this dynamical balance, without requiring user input. The method is able to achieve low numerical dissipation in absence of oscillations or diminish oscillation on skew meshes, while it shows minimal loss in accuracy for a turbulent test case. Despite its advantages, the method exhibits a slight decrease in the second-order relation between time-step size and pressure error, suggesting that other feedback mechanisms could be of interest.
\end{abstract}

\begin{keyword}
checkerboarding \sep collocated grids \sep conservative discretisation
\end{keyword}

\end{frontmatter}


\section{Introduction}
\label{sec:Introduction}
\FloatBarrier

\begin{figure}[ht]
    \centering
    \includegraphics[width=0.8\textwidth]{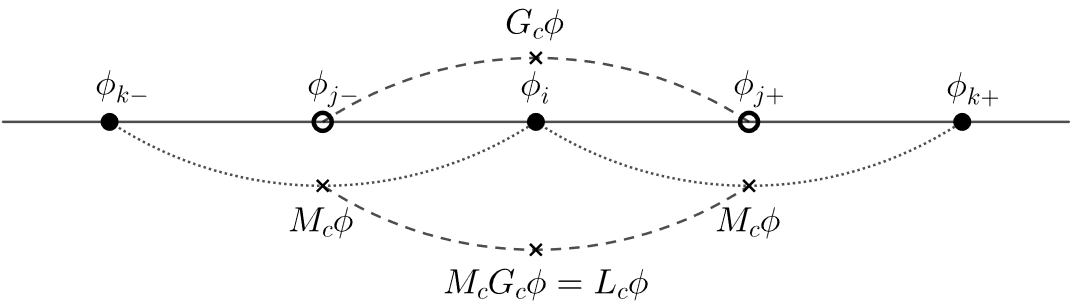}
    \caption{Schematic drawing of how the Central Difference Scheme leads to decoupling of nodes.}
    \label{fig:CDS}
\end{figure}

The checkerboard problem arises due to the non-trivial pressure-velocity coupling for incompressible Newtonian fluid flows. The governing equations for such flows are given by the momentum and continuity equations:

\begin{align}
    \dt\uvec + \br{\uvec\cdot\nabla} \uvec &= \nu\nabla^2\uvec - \frac{1}{\rho}\nabla p, \label{eq:NSMomCont} \\
    \nabla\cdot\uvec &= 0, \label{eq:NSContCont}
\end{align}

\noindent which, for three spatial dimensions, provide only three independent equations and four unknowns. For incompressible flows, the pressure cannot be obtained from the equation of state, and formulating an equation for pressure becomes non-trivial, a topic that has been widely studied in the field of computational fluid dynamics (CFD) \cite{Patankar1980, Wesseling2009, Ferziger2002, Versteeg2007}. Many methods solve this problem iteratively by predicting a velocity field and then finding a pressure field of which the gradient projects it onto a divergence-free space, using the Helmholtz-Hodge theorem \cite{Chorin1968, Temam1969, Patankar1983, Kim1985, VanKan1986}. If the discrete gradient at node $i$ is derived through central differencing, its value will only depend on the pressure at neighbouring nodes $j$ and not on the pressure at node $i$ itself. Moreover, if done consistently, the discrete Laplacian operator derived from this gradient will be based on a wide stencil. In this stencil, node $i$ is coupled to nodes $k$, which are neighbours to nodes $j$, resulting in a decoupling between node $i$ and directly neighbouring nodes $j$, see figure \ref{fig:CDS}. The use of this so-called wide-stencil Laplacian in the Poisson equation and the aforementioned gradient in the velocity correction results in a decoupling of the pressure field between neighbouring cells. This in turn can lead to non-physical oscillations and checkerboard-like patterns in the pressure field, also known as the checkerboard problem \cite{Patankar1980}.

On structured Cartesian meshes, this problem can be circumvented by using a staggered grid arrangement in which the velocities at the cell faces are coupled to a compact-stencil gradient of pressure between directly neighbouring cell-centered nodes \cite{Harlow1965}. However, for many industrial applications the use of CFD involves complex geometries that require unstructured meshes. Extension of the staggered grid method to such cases is not straight-forward and leads to complex correction schemes and increased computational costs \cite{Perot2000, Pascau2011}. 

One commonly used strategy to address pressure field oscillations, is the use of the weighted interpolation method (WIM) to evaluate the velocity at the cell faces, in contrast to the direct interpolation method (DIM). This method was originally introduced as the pressure-weighted interpolation method and has allowed the wide-spread usage of the collocated grid arrangement \cite{Rhie1983, Peric1988}. Since its introduction this method has been extended in many ways, e.g. to account for decoupling caused by the use of small time-steps found in transient flows \cite{WenZhongShen2001, Choi1999, Kawaguchi2002, yu2002discussion} and to account for under-relaxation factors \cite{Majumdar1988, Miller1988, Yu2002}. Generalisation and unification of the aforementioned problems and accompanying solutions can be found in more recent works \cite{Guermond2006, Zhang2014, Bartholomew2018}, which offer a comprehensible overview of these methods. One unavoidable consequence that these solutions have in common is the introduction of a numerical error in the form of a non-zero contribution to the kinetic energy balance of either the pressure term or the convective term \cite{Felten2006}. Nevertheless, many commercially available and open-source codes favour stability at the cost of accuracy \cite{Ansys2023, STARCCM2023, greenshields2023, archambeau2004code}, by applying a dissipative form of the WIM. This avoids pressure field oscillations while at the same time increasing stability and accepting the consequential numerical error. Usually this is done implicitly through the use of a compact-stencil Laplacian, which directly couples node $i$ to its neighbouring nodes $j$ instead of its second-neighbours $k$. In this method, divergence-free values for the velocities at the faces follow directly from the Poisson equation, without introducing any numerical error. However, the coupling effect of the WIM is introduced implicitly by the Laplace operator and the numerical error is introduced to the collocated velocities in this case. This method introduces a dissipative pressure error, with the benefit of greatly reducing computational complexity and cost when using unstructured grids. 

However, with the increase in computational resources available and the ensuing rise of high-fidelity simulations, higher accuracy is desired to describe motion of fluids. Numerical dissipation is limiting this accuracy by disrupting fluid motion, especially at the smaller scales, which are essential to accurately depict turbulent flows, making the use of conservative schemes more important. The symmetry-preserving method eliminates numerical dissipation and conserves the physical properties of the flow, while at the same time warranting unconditional stability, by mimicking the properties of the continuous operators in their discrete counterparts \cite{Verstappen2003}. This method has been extended to collocated grid arrangements \cite{Trias2014} and implemented in the open-source code OpenFOAM \cite{Komen2021}. The pressure error that is introduced by the application of this method on collocated grids remains as the largest source of numerical dissipation. 

Other methods to filter pressure oscillations without the introduction of numerical dissipation have been attempted, such as  filtering pressure modes that lie on the kernel of the wide-stencil Laplacian operator \cite{larsson2010}. However, through examination of the connection between the mesh and this kernel, it was shown that this method only works on Cartesian meshes \cite{Hopman2023d}, as the oscillatory part of the kernel vanishes for most unstructured meshes and complex geometries. A method that can eliminate, or at least balance, both the checkerboard problem and numerical dissipation is therefore still sought after. Moreover, the inadequacy of the kernel filtering method illustrated that a broader and clear definition of checkerboarding is lacking, and most published works in literature use a qualitative description of the phenomenon.  

The present work contributes to this topic by introducing a clear definition of the checkerboard problem. This definition can be used to quantify the level of checkerboarding on any geometry or mesh. Furthermore, a solver algorithm is developed which uses this quantification method to dynamically balance the occurrence of pressure oscillations and numerical dissipation. The structure of this paper is as follows: Section \ref{sec:numFramework} gives an overview of the equations and different algorithms that are used and discusses the occurrence of checkerboarding and numerical errors. Section \ref{sec:quantifyCB} discusses methods to quantify checkerboarding and introduces a solver that dynamically balances numerical dissipation and checkerboard oscillations. Section \ref{sec:results} shows results for the new solver compared to existing methods. Finally, section \ref{sec:conclusions} discusses the conclusions of the work and the outlook for future work.

\FloatBarrier

\section{Numerical framework}
\label{sec:numFramework}
\subsection{Symmetry-preserving method}
\label{sec:spmeth}
\FloatBarrier

The semi-discretised formulation of equations \eqref{eq:NSMomCont} and \eqref{eq:NSContCont} for arbitrary collocated grids, using the matrix-vector notation of \cite{Trias2014}, is given by:

\begin{align}
    \Omega \dt\uvec_c + \C\uvec_c &= -D\uvec_c - \Omega G_c\pvec_c \label{eq:NSMomDis}, \\
    M\uvec_s &= \mathbf{0}_c \label{eq:NSContDis}.
\end{align}

\noindent In three dimensions, the discrete collocated kinematic pressure and velocity fields are given by $\pvec_c\in\mathbb{R}^n$ and $\uvec_c = \br{\uvec_{c,x}^T, \uvec_{c,y}^T, \uvec_{c,z}^T}^T\in\mathbb{R}^{3n}$, whereas $\uvec_s\in\mathbb{R}^m$ gives the staggered velocities, in which $n$ and $m$ give the number of control volumes and faces respectively. A small set of matrices containing geometric information about the mesh, given in table \ref{tab:SPBuildBlocks}, is enough to derive all the matrix operators needed to from an algebraic symmetry-preserving framework, given in table \ref{tab:SPOperators}.

\begin{figure}[ht]
    \centering
    \includegraphics[width=0.5\textwidth]{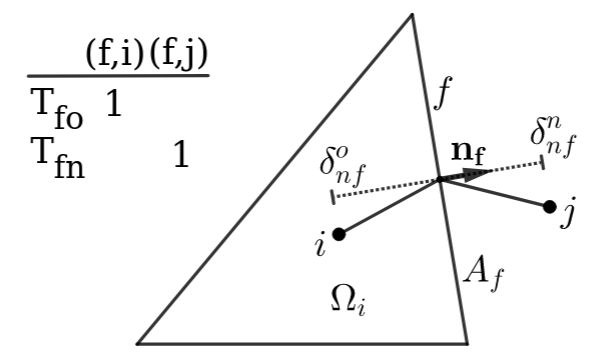}
    \caption{Geometric parameters of the operators in table \ref{tab:SPBuildBlocks}, needed to establish the symmetry-preserving scheme.}
    \label{fig:geom}
\end{figure}

\begin{table}[ht]
\centering
\begin{tabular}{llp{0.63\linewidth}}
    \toprule
    Operator & Dimensions & Description \\
    \midrule
    $T_{fo}, T_{fn}$ & $n\times m$ & face-owner and face-neighbour connectivity matrices, containing entry $(i,f) = 1$ if cell $i$ is connected to face $f$ as an owner or as a neighbour respectively. \\ 
    \midrule
    $N_s$ & $m\times 3m$ & face-normal matrix with diagonal blocks $\br{ N_{sx}, N_{sy}, N_{sz}}$ containing the $\br{x,y,z}$-components of the face-normals, $\mathbf{n}_f$, respectively. \\ 
    \midrule
    $A_s$ & $m\times m$ & diagonal matrix containing face areas, $A_f$. \\ 
    \midrule
    $\delta_{ns}^o, \delta_{ns}^n$ & $m\times m$ &  face-owner and face-neighbour normal distance diagonal matrices containing $\delta_{nf}^o$ and $\delta_{nf}^n$ respectively, denoting the absolute values of the $\mathbf{n}_f$-projected vectors from face-centroid to owner or neighbour centroid respectively. \\ 
    \midrule
    $\Omega_c$ & $n\times n$ & cell-volume diagonal matrix. \\
    \bottomrule
\end{tabular}
\caption{Matrices containing the geometric parameters shown in figure \ref{fig:geom}, which form the building blocks for all symmetry-preserving operators. $m$ and $n$ denote the number of faces and control volumes respectively.}    
\label{tab:SPBuildBlocks}
\end{table}

\begin{table}[ht]
\centering
\begin{tabular}{llp{0.43\linewidth}}
    \toprule
    Operator definition & Dimensions & Description \\
    \midrule
    $\Omega = I_3\otimes\Omega_c$ & $3n\times3n$ & collocated volumes \\
    \midrule
    $\delta_{ns} = \delta_{ns}^o + \delta_{ns}^n$ & $m\times m$ & face-normal distances \\
    \midrule
    $\Omega_s = \delta_{ns}A_s$ & $m\times m$ & staggered volumes \\
    \midrule
    $S_s = A_sN_s$ & $m\times 3m$ & surface vectors \\
    \midrule
    $M = \br{ T_{fo} - T_{fn}} A_s$ & $n\times m$ & divergence \\
    \midrule
    $\begin{aligned}
        G &= \delta_f^{-1}\br{ T_{fn} - T_{fo}}^T \\
        &= -\Omega_s^{-1}M^T
    \end{aligned}$ & $m\times n$ & gradient \\
    \midrule
    $L = MG = -M\Omega_s^{-1}M^T$ & $n\times n$ & compact-stencil Laplacian \\
    \midrule
    $W_o^{\gamma} =
    \begin{cases}
        \delta_{ns}^{-1}\delta_{ns}^n, & \gamma=L \\
        \frac{1}{2}I_m, & \gamma=M \\
        \delta_{ns}^{-1}\delta_{ns}^o, & \gamma=V
    \end{cases}$ & $m\times m$ & interpolation weights\\
    \midrule
    $W_n^{\gamma} = I_m - W_o^{\gamma}$ \\
    \midrule
    $\Pi_{cs}^{\gamma} = W_o^{\gamma}T_{fo}^T + W_n^{\gamma}T_{fn}^T$ & $m\times n$ & cell-to-face interpolator \\
    \midrule
    $\Gamma_{cs}^{\gamma} = N_s\br{ I_3\otimes\Pi_{cs}^{\gamma}}$ & $m\times 3n$ & cell-to-face dot-interpolator \\
    \midrule
    $\Gamma_{sc}^{\gamma} = \Omega^{-1}\Gamma_{cs}^{\gamma^T}\Omega_s$ & $3n\times m$ & face-to-cell interpolator\\
    \midrule
    $M_c^{\gamma} = M\Gamma_{cs}^{\gamma}$ & $n\times 3n$ & collocated divergence \\
    \midrule
    $G_c^{\gamma} = \Gamma_{sc}^{\gamma}G = -\Omega^{-1}M_c^{\gamma^T}$ & $3n\times n$ & collocated gradient \\
    \midrule
    $\begin{aligned}
        L_c^{\gamma} &= M_c^{\gamma}G_c^{\gamma} \\
        &= -M\Gamma_{cs}^{\gamma}\Omega^{-1}\Gamma_{cs}^{\gamma^T}M^T
    \end{aligned}$ & $n\times n$ & wide-stencil Laplacian \\
    \midrule
    $\C[c][] = M\text{diag}\br{\uvec_s}\Pi_{cs}^M$ & $n \times n$ & convective block\\    
    \midrule
    $\C = I_3\otimes \C[c][]$ & $3n \times 3n$ & convective operator \\
    \midrule
    $D_c = -\nu L$ & $n\times n$ & diffusive block \\
    \midrule
    $D = I_3\otimes D_c$ & $3n\times 3n$ & diffusive operator \\
    \bottomrule
\end{tabular}
\caption{Full set of matrix operators to form an algebraic symmetry-preserving framework. $m$ and $n$ denote the number of faces and control volumes respectively. $\gamma$ indicates an unspecified interpolation method, for which $L, M \text{ and } V$ are given as options, indicating \emph{linear}, \emph{midpoint} and \emph{volumetric} interpolation respectively.}
\label{tab:SPOperators}
\end{table}

\noindent Midpoint interpolation in the convective term is necessary to maintain the skew-symmetry of the continuous operator \cite{Verstappen2003}, whereas it was shown in \cite{Hopman2022a, Santos2022, Santos2023} that employing volumetric interpolation in the other operators leads to unconditional stability, even on highly-distorted meshes. In the following text therefore, if superscript $\gamma$ is dropped, volumetric interpolation is employed, e.g. $L_c = M_cG_c = -M\Gamma_{cs}\Omega^{-1}\Gamma_{cs}^TM^T = -M\Gamma_{cs}^V\Omega^{-1}\Gamma_{cs}^{V^T}M^T$.

\FloatBarrier

\subsection{Time-stepping algorithm}
The temporal integration, which was left undiscretised in equation \eqref{eq:NSMomDis}, is taken care of by the fractional step method \cite{Yanenko1971}, in which a predictor velocity is calculated and corrected through solving a Poisson equation for pressure. Usage of a wide-stencil Laplacian for the pressure Poisson equation in combination with the DIM to calculate the staggered velocities leads to checkerboarding. By applying a compact stencil and/or the WIM, this method can be adjusted to deal with pressure field oscillations. Table \ref{tab:FSM} gives an overview of the four possible fractional step methods found by combining the compact- and wide-stencil methods with the DIM and the WIM. The temporal discretisation is not the main interest in this work and is simply denoted by a function, $\mathcal{F}\br{\uvec_c, \uvec_s}$, which calculates the velocity predictor, $\uvec_c^{p}$. As an example, for Forward Euler time-stepping, this term is given by: 

\begin{equation}
\begin{split}
\label{eq:Rucus}
    \uvec_c^{p} &= \mathcal{F}\br{\uvec_c, \uvec_s}, \\
                &= \uvec_c^n + \Delta tR\br{\uvec_c^n,\uvec_s^n}, \\
                &= \uvec_c^n - \Delta t\Omega^{-1}\br{ \C[][][n] + D}\uvec_c^n.
\end{split}
\end{equation}

\begin{table}[ht]
\centering
\begin{tabular}{C{0.2\linewidth} c C{0.2\linewidth} c C{0.2\linewidth} c C{0.2\linewidth}}
    \multicolumn{3}{c}{Wide stencil} & & \multicolumn{3}{c}{Compact stencil} \\
    DIM & & WIM & & DIM & & WIM \\
    \toprule
    \multicolumn{7}{c}{\cellcolor{C1}$\uvec_c^{p} = \mathcal{F}\br{\uvec_c,\uvec_s}$} \\
    \\[-2ex]    
    \multicolumn{3}{c}{\cellcolor{C1}} & & \multicolumn{3}{c}{\cellcolor{C1}$\uvec_c^{p*}=\uvec_c^{p}-G_c\ptil_c^p$} \\
    \\[-2ex]
    \multicolumn{3}{c}{\cellcolor{C1}$L_c\ptil_c^{n+1}=M_c\uvec_c^{p}$} & & \multicolumn{3}{c}{\cellcolor{C1}$L\ptil_c' = M_c\uvec_c^{p*}$} \\
    \\[-2ex]
    \multicolumn{3}{c}{\cellcolor{C1}} & & \multicolumn{3}{c}{\cellcolor{C1}$\ptil_c^{n+1} = \ptil_c^p + \ptil_c'$} \\
    \\[-2ex]
    \multicolumn{3}{c}{\cellcolor{C1}$\uvec_c^{n+1} = \uvec_c^{p} - G_c\ptil_c^{n+1}$} & & \multicolumn{3}{c}{\cellcolor{C1}$\uvec_s^{n+1} = \Gamma_{cs}\uvec_c^{p}-G\ptil_c^{n+1}$} \\
    \\[-2ex]    
    \cellcolor{C1}$\uvec_s^{n+1} = \Gamma_{cs}\uvec_c^{n+1}$ & & \cellcolor{C1}$\uvec_s^{n+1} = \Gamma_{cs}\uvec_c^{p} - G\ptil_c^{n+1}$ & & \cellcolor{C1}$\uvec_c^{n+1} = \Gamma_{sc}\uvec_s^{n+1}$ & & \cellcolor{C1}$\uvec_c^{n+1} = \uvec_c^{p} - G_c\ptil_c^{n+1}$ \\
\end{tabular}
\caption{Overview of the four possible fractional step methods found by combining the compact- and wide-stencil methods with the DIM and the WIM.}
\label{tab:FSM}
\end{table}

The value for the pressure predictor, $\ptil_c^p$ is usually chosen to be $\mathbf{0}_c$ or $\ptil_c^n$, corresponding to the classical Chorin projection method \cite{Chorin1967} and  the second-order Van Kan projection method \cite{VanKan1986}, respectively. Note that this pressure predictor only has an effect when a compact-stencil Laplacian is used. For the wide-stencil methods, $L_c\ptil_c^p$ could simply be added to both sides of the Poisson equation, resulting in $L_c\ptil_c^{n+1}$ on the left hand side (LHS) and $M\Gamma_{cs}\uvec_c^{p}$ on the right hand side (RHS).

\noindent Upon closer inspection, the compact-stencil DIM can immediately be discarded for being less accurate than the compact-stencil WIM. This is true because the only difference between these methods is the back-and-forth interpolation of the predictor velocity. This operation can be viewed as the application of a Laplacian filter, as: $\uvec_c^{n+1} = \Gamma_{sc}\Gamma_{cs}\uvec_c^p - G_c\ptil_c' = \mathcal{L}_f\br{\uvec_c^p} - G_c\ptil_c'$. This filter can be understood most easily by considering its effect on a uniform Cartesian mesh, where the classical $\sbr{1, -2, 1}$ coefficients of the Laplacian are retrieved: 

\begin{align}
    \mathcal{L}_f\br{\uvec_c^p} &= L_f\uvec_c^p \qquad \textit{(uniform Cartesian)}, \\
    L_f &= I + \frac{1}{4}\text{diag}\br{ L_{f,x}, L_{f,y}, L_{f,z}}, \\
    \sbr{L_{f,x}\uvec_{c,x}^p}_{i,j,k} &= \sbr{\uvec_{c,x}^p}_{i-1,j,k} - 2\sbr{\uvec_{c,x}^p}_i + \sbr{\uvec_{c,x}^p}_{i+1,j,k},
\end{align}

\noindent where subscripts $i,j,k$ indicate cell numbering in $x,y,z$-directions respectively. This unnecessary extra filtering is smoothing the velocity field and thereby causing undesired numerical dissipation. This method is therefore not considered any further.

The remaining methods each have their own problem, which makes choosing the right method a trade-off between these factors. As discussed in the introduction, the wide-stencil DIM is very prone to the occurrence of checkerboarding. In the wide-stencil WIM, a correction is applied to the staggered velocities, which makes them divergent as a result. Although carried out in exactly the same way, the calculation of the staggered velocities for the compact-stencil WIM is without correction, and follows directly from the Poisson equation. In this method, however, the correction is applied to the collocated velocities, which in turn makes them non-divergence-free. The divergence of either the staggered or collocated velocities introduces a numerical error to the simulation, which is further discussed in section \ref{sec:numdiss}.

\subsection{Occurrence of checkerboarding}
\label{sec:occurrence}
Although the WIM, for both the compact- and wide-stencil methods, introduces a coupling of the pressure field between neighbouring nodes at the cost of a numerical error, these methods can still show oscillating pressure fields, most commonly caused by the usage of a small time-step in unsteady simulations \cite{Felten2006}, or by the inclusion of a pressure predictor as $\ptil_c^p = \ptil_c^n$ in case of the compact-stencil WIM \cite{Komen2021}. To illustrate this, note that if $\Delta t \rightarrow 0^+$, then $\uvec_c^{p} \rightarrow \uvec_c^n$ in equation \eqref{eq:Rucus}, since the effect of the convective and diffusive terms are proportional to $\Delta t$. This removes the coupling established in the wide-stencil WIM, since it is dependent on $\uvec_s$ and the convective term. To show the decoupling for the compact-stencil WIM, first regard the case of $\ptil_c^p = \mathbf{0}_c$, with Forward Euler temporal discretisation as an example, which leads to:

\begin{align}
    \uvec_c^{p*} &= \uvec_c^n, \\
    \uvec_c^n &= \uvec_c^{n-1} - G_c\ptil_c^n = \uvec_c^0 - G_c\sum_i^n\ptil_c^i, \\
    L\ptil_c^{n+1} &= M_c\uvec_c^{p*} = M_c\uvec_c^0 - L_c\sum_i^n\ptil_c^i, \label{eq:compPoisDec} \\
    L\mathbb{P}_c^{n+1} &= M_c\uvec_c^0 + \br{L - L_c}\mathbb{P}_c^n,
    \label{eq:compPoisStatIt}
\end{align}

\noindent where $\mathbb{P}^n = \sum_i^n\Tilde{\mathbf{p}}_c^i$ and $\mathbb{P}^n$ is added on both sides of equation \eqref{eq:compPoisDec} to reach equation \eqref{eq:compPoisStatIt}. Equation \eqref{eq:compPoisStatIt} gives a stationary iterative method to solve the wide-stencil Poisson equation, to which the solution is a decoupled pressure field. The coupling that the compact-stencil Laplacian gave is therefore lost if the time-step becomes too small. Using the Van Kan method, where $\ptil_c^p = \ptil_c^n$, further accelerates the problem. In this case the wide-stencil Laplacian operates on a larger part of the pressure field:

\begin{equation}
    L\ptil_c' = M_c\uvec_c^p = M_c\uvec_c^{p*} - L_c\ptil_c^p,
\end{equation}

\noindent which weakens the coupling that the compact-stencil Laplacian provided. 

Since the pressure oscillations are "invisible" to the wide-stencil gradient, the collocated velocities remain unaffected and the algorithm advances as if they were not there. The oscillations are retained in this case and could grow until they reach a point in which they cause numerical issues and unstable solutions. Although their retention is evident, their actual origins and method of growth are not entirely understood. In \cite{Trias2014} it was argued that the convective term causes spurious modes in the velocity field that lead to checkerboarding, but an analysis of this process was not performed. A relation to the method with which the Poisson equation is solved, including preconditioners, was discussed in \cite{Hopman2023}, but the exact mechanisms remain unclear.

\subsection{Numerical errors of the WIM}
\label{sec:numdiss}
To analyse the numerical errors of the WIM, the global discrete kinetic energy is considered, given by $E_K = \frac{1}{2}\uvec_c^T\Omega\uvec_c$. The temporal evolution of this term is derived using the product rule and equation \eqref{eq:NSMomDis}:

\begin{equation}
\label{eq:tEvoEKin}
    \frac{d}{dt}E_K = -\frac{1}{2}
    \begin{pmatrix*}[l]
        \phantom{+}\uvec_c^T\br{ \C + \C[][T]}\uvec_c \\
        +\uvec_c^T\br{ D + D^T}\uvec_c \\
        +\uvec_c^T\Omega G_c \pvec_c + \pvec_c^TG_c^T\Omega^T\uvec_c
    \end{pmatrix*}.
\end{equation}

\noindent If $\C$ is symmetry-preserving, i.e. $\C=-\C[][T]$, mimicking the continuous operator, then the contribution of the convective term equals zero. Similarly, the contribution of the pressure term equals zero if $-M_c^T = \Omega G_c$ and $M_c\uvec_c = \mathbf{0}_c$. The global kinetic energy dissipation is therefore reduced to the viscous term, equaling:

\begin{equation}
    \frac{d}{dt}E_K = -\frac{1}{2}\uvec_c^T\br{ D + D^T}\uvec_c \leq 0,
\end{equation}

\noindent which is strictly dissipative if $D$ is constructed using the symmetry-preserving discretisation, retaining the positive-definiteness of the continuous operator.

\subsubsection{Wide-stencil WIM - Convective error}
\label{sec:numdiss_wsWIM}
For the wide-stencil WIM, $M_c\uvec_c^{n+1} = \mathbf{0}_c$ and therefore the pressure error equals zero. Conversely, the convective error is non-zero. This is caused by the correction to the staggered velocities and the resulting fact that $M\uvec_s^{n+1} \neq 0$, since it is given by:

\begin{equation}
\begin{split}
    M\uvec_s^{n+1} &= M\br{\Gamma_{cs}\uvec_c^p-G\ptil_c^{n+1}}, \\
    &= \cancelto{\mathbf{0}_c}{M\Gamma_{cs}\br{\uvec_c^p -G_c\ptil}} + M\br{\Gamma_{cs}\Gamma_{sc} - I} G\ptil_c^{n+1}, \\
    &= \br{ L_c-L}\ptil_c^{n+1},
\end{split}
\end{equation}

\noindent To see the effect on the evolution of kinetic energy, recall that $\C[c] =  M\text{diag}\br{\uvec_s}\Pi_{cs}^m$. The off-diagonals of $\C[c]$ contain the fluxes between cells, and are skew-symmetric by construction. In contrast the diagonal entries are given by:

\begin{equation}
    \C[i,i] = \frac{1}{2}\sbr{M\uvec_s}_i,
\end{equation}

\noindent which are non-zero in this case. If $\C[c]$ is split into its non-zero diagonal and skew-symmetric off-diagonals as $\C[c] = \text{diag}\br{M\uvec_s} + \C[c][OD]$, then the numerical error of the evolution of the global kinetic energy at time-step $n+1$ due to the convective error is given by:

\begin{equation}
\label{eq:convErr}
\begin{split}
    \frac{d}{dt}\mathcal{E}_K = 
    &-\frac{1}{2} \uvec_c^{n+1^T}\br{\cancelto{\mathbf{0}^{3n\times 3n}}{\C[][OD][n+1] + \C[][OD][n+1]}}\uvec_c^{n+1}, \\
    &-\frac{1}{2} \uvec_c^{n+1^T}\sbr{I_3\otimes\text{diag}\br{M\uvec_s^{n+1}}}\uvec_c^{n+1}.
\end{split}
\end{equation}

\noindent The convective error in the wide-stencil WIM is therefore caused by the divergence of the staggered velocities and proportional to $\Delta t\br{L_c-L}\pvec_c^{n+1}$.

\subsubsection{Compact-stencil WIM - Pressure error}
\label{sec:numdiss_csWIM}
On the other hand, for the compact-stencil WIM, $M\uvec_s^{n+1} = \mathbf{0}_c$ and therefore the convective error equals zero. However, in this case the correction of the collocated velocities leads to $M_c\uvec_c^{n+1} \neq \mathbf{0}_c$, which is given by:

\begin{equation}
\begin{split}
    M_c\uvec_c^{n+1} &= M\Gamma_{cs}\br{\uvec_c^p-G_c\ptil_c'}, \\
    &= \cancelto{\mathbf{0}_c}{M\br{\Gamma_{cs}\uvec_c^p}-G\ptil_c'} + M\br{I-\Gamma_{cs}\Gamma_{sc}}G\ptil_c', \\
    &= \br{L-L_c}\ptil_c'.
\end{split}
\end{equation}

\noindent Then the ensuing numerical error in the evolution of the global kinetic energy at time-step $n+1$ due to the pressure error is given by:

\begin{equation}
\label{eq:presErr}
\begin{split}
    \frac{d}{dt}\mathcal{E}_K &= -\uvec_c^{n+1^T}\Omega G_c\pvec_c', \\
    &= \pvec_c'^TM_c\uvec_c^{n+1}, \\
    &= \pvec_c'^T\br{L-L_c}\ptil_c'.
\end{split}    
\end{equation}

\noindent In the compact-stencil WIM, the pressure error is caused by the divergence of the collocated velocities and is, similarly to the convective error, proportional to $\Delta t\br{L-L_c}\pvec_c'$. 

In practice, most numerical codes revert to the compact-stencil WIM. The main reasons for this are threefold: (i) The inclusion of a pressure predictor as $\ptil_c^p = \ptil_c^n$ reduces the order of the pressure error from $\mathcal{O}\br{\Delta t^2}$ to $\mathcal{O}\br{\Delta t^4}$ since $\ptil_c^{n+1} \sim \mathcal{O}\br{\Delta t}$ and $\br{\ptil_c^{n+1}-\ptil_c^n} \sim \mathcal{O}\br{\Delta t^2}$, making the non-physical contribution quite small. This reduction is not possible with the wide-stencil Laplacian. (ii) The compact-stencil Laplacian reduces the computational complexity and cost when using unstructured grids. And finally, (iii) the convective error term, as shown in equation \eqref{eq:convErr}, is not strictly dissipative. Since the global divergence of the staggered velocities will always be equal to zero, there will always be cells with negative divergence to compensate cells that have positive divergence. This means that kinetic energy can be added in some parts of the solution, leading to instabilities. Conversely, the resulting term of equation \eqref{eq:presErr} will be dependent on the term $\br{I-\Gamma_{cs}\Gamma_{sc}}$, which in turn depends on the choice in interpolators and meshing. These factors are much easier to control, and moreover, it was shown in \cite{Hopman2022a, Santos2022, Santos2023} that choosing volumetric interpolation for this operator causes this term to be strictly dissipative, even on highly-distorted meshes, warranting stability.

\section{Defining the checkerboard problem}
\label{sec:quantifyCB}
Past works that extensively discuss the topic use different names for the checkerboard problem, such as checkerboarding \cite{Larmaei2010, Kawaguchi2002, Klaij2015}, spurious pressure modes \cite{Guermond2006}, odd-even decoupling \cite{Rauwoens2007}, or zigzagness \cite{Date2003}, but none of them give a quantifiable definition for the problem. In those works a constant and uniform solution is applied that does not need quantification of the problem, whereas in this work a solution is sought after that is applied proportionally to the severity of the problem. For this reason a clear quantification method has to be introduced first.

\subsection{Using the kernel of the discrete Laplacian to define checkerboarding}
One way to define checkerboard modes is by performing an eigenvector decomposition of $L_c^{\gamma}$ and defining the checkerboard modes as $\mathbf{p}_c^-\in Ker(L_c^{\gamma})$. Since $L_c^{\gamma} = M_c^{\gamma}G_c^{\gamma} = M\Gamma_{cs}^{\gamma}\Gamma_{sc}^{\gamma}G$, the $\mathbf{p}_c^-$ modes will depend on the choice of interpolator and the mesh. The constant mode vector will always lie on the kernel, in addition to any spurious mode vectors. This method was for instance used by \cite{larsson2010}, where the method was adequate since midpoint interpolation and Cartesian meshes were used. By looking at the definitions of $M_c^{\gamma}$ and $G_c^{\gamma}$ it becomes clear why this null-space exists for these specific conditions, since:

\begin{equation}
\begin{split}
\label{eq:midpointGrad}
    \sbr{\Gamma^{\gamma}_{sc}G\pmb{\phi}_c}_i &= \frac{1}{\Omega_i}\sum_{f\in F_f(i)}w_{fi}^{\gamma}\Omega_f\frac{\pmb{\phi}_j - \pmb{\phi}_i}{\delta_{nf}}\mathbf{n}_{f(i)}, \\
    &= \frac{1}{\Omega_i}\sum_{f\in F_f(i)}w_{fi}^{\gamma}\pmb{\phi}_j\mathbf{s}_{f(i)} - \frac{1}{\Omega_i}\sum_{f\in F_f(i)}w_{fi}^{\gamma}\pmb{\phi}_i\mathbf{s}_{f(i)},
\end{split}
\end{equation}

\noindent gives the wide-stencil gradient at cell $i$. Where $\pmb{\phi}_c \in \mathbb{R}^{n\times 1}$ and $\pmb{\phi}_i = \sbr{\pmb{\phi}_c}_i$. Cell $j$ is neighbouring cell $i$ through face $f$. $F_f(i)$ denotes the set of faces that define cell $i$. $\Omega_i = \sbr{\Omega_c}_{i,i}$, $\Omega_f = \sbr{\Omega_s}_{f,f}$ and $\delta_{nf} = \sbr{\delta_{ns}}_f$. $w_{if}$ equals $\sbr{W_o^{\gamma}}_f$ or $\sbr{W_o^{\gamma}}_f$ depending on whether $i$ is the owner or the neighbour of face $f$, respectively. Finally, $\mathbf{n}_{f(i)}$ and $\mathbf{s}_{f(i)}$ respectively give the outward-pointing face-normal vector and outward-pointing surface vector at face $f$ with respect to cell $i$, note that $\mathbf{s}_{f(i)} = A_f \mathbf{n}_{f(i)}$. The second term of the RHS of equation \eqref{eq:midpointGrad} goes to zero in case $w_{fi}^{\gamma}$ is constant over all faces of $i$, since $\sum_{f\in F_f(i)}\mathbf{s}_{f(i)} = \mathbf{0}$. This is the case for midpoint interpolation and for uniform Cartesian meshes. Similarly for the divergence operator:

\begin{equation}
\begin{split}
\label{eq:midpointDiv}
    \sbr{M\Gamma_{cs}^{\gamma}\pmb{\psi}_c}_i &= \sum_{f\in F_f(i)}\br{w_{if}^{\gamma}\pmb{\psi}_i + w_{jf}^{\gamma}\pmb{\psi}_j}\cdot\mathbf{s}_{f(i)}, \\
    &=\sum_{f\in F_f(i)}w_{if}^{\gamma}\pmb{\psi}_i\cdot\mathbf{s}_{f(i)} + \sum_{f\in F_f(i)}w_{jf}^{\gamma}\pmb{\psi}_j\cdot\mathbf{s}_{f(i)},
\end{split}
\end{equation}

\noindent in which the first term of the RHS goes to zero for midpoint interpolation and for uniform Cartesian meshes. When combining equations \eqref{eq:midpointGrad} and \eqref{eq:midpointDiv} it becomes evident that $L_c^{\gamma}$ does not connect cell $i$ to its direct neighbours $j$, but only to its second-neighbours $k$, when midpoint interpolation is used:

\begin{equation}
\label{eq:LapEntries}
    \sbr{L_c^M}_{i,j} = 0, \quad \sbr{L_c^M}_{i,k} = \frac{A_fA_g}{4\Omega_j} \mathbf{s}_{f(i)}\cdot\mathbf{s}_{g(j)},
\end{equation}

\noindent in which face $f$ lies between cells $i$ and $j$, and face $g$ lies between cells $j$ and $k$. If this odd-even parity is sustained throughout the entire mesh, two disconnected groups of cells will exist. In addition to the constant kernel vector, this gives rise to a spurious kernel vector:

\begin{equation}
    \sbr{\mathbf{p}_c^-}_i = (-1)^{p(i)},
\end{equation}

\noindent where $p(i)$ denotes the parity of cell $i$, equal to 0 or 1. In the special case of Cartesian meshes, the dot product in equation \eqref{eq:LapEntries} will be zero for any diagonal second-neighbour pairing, giving rise to $2^{N_{dim}}$ kernel vectors, with number of dimensions $N_{dim}$. For example, in three-dimensional Cartesian meshes the resulting set of eight vectors is given by \cite{larsson2010}:

\begin{equation}
    \sbr{\mathbf{p}_{c(IJK)}^-}_{i,j,k} = (-1)^{iI+jJ+kK},
\end{equation}

\noindent in which $I,J,K \in \{0,1\}$ and indices $i,j,k$ indicate the cell numbering in each of the Cartesian directions. In this notation, $\mathbf{p}_{c(000)}^-$ gives the constant vector. $\sbr{L_c^{\gamma}}_{i,j}$ is generally only zero for midpoint interpolation. Whereas, in general, $\sbr{L_c^{\gamma}}_{i,j}\neq 0$ for linear or volumetric interpolation, except on uniform meshes where all interpolators equal midpoint interpolation. However, a set of kernel vectors was derived by \cite{Hopman2023d} for any Cartesian mesh with midpoint, linear or volumetric interpolation. To do so, the fact that $Ker(G_c^{\gamma}) \in Ker(L_c^{\gamma})$ was used, in addition to rewriting $G_c^{\gamma}$ to $G_G\overline{\Pi}_{cs}^{\gamma}$. $G_G$ denotes a Gauss-gradient and the overbar denotes swapping the interpolation weights between owners and neighbours of the faces, such that $\Pi_{cs}^M = \overline{\Pi}_{cs}^M$, $\Pi_{cs}^V = \overline{\Pi}_{cs}^L$ and $\Pi_{cs}^L = \overline{\Pi}_{cs}^V$. For details on this equality, the reader is referred to \ref{sec:ApGc}. By taking the interpolation before the Gauss-gradient, cell-centered values can be chosen such that sets of opposing interpolated face values are always equal, for which the cell-centered Gaus-gradient equals zero. The resulting set of eight vectors for three-dimensional Cartesian meshes is then given by:

\begin{equation}
\label{eq:KerVecGeneral}
    \sbr{\mathbf{p}_{c(IJK,\gamma)}^-}_{i,j,k} = (-1)^{iI+jJ+kK}\br{\sbr{\Delta x}_i^I\sbr{\Delta y}_j^J\sbr{\Delta z}_k^K}^{\alpha},
\end{equation}

\noindent in which $\alpha = \{-1, 0, 1\}$ for linear, midpoint and volumetric interpolations respectively. This set is not necessarily mutually orthogonal, especially for linear and volumetric interpolations, however, in all cases the set is linearly independent and therefore spans the null-space of $L_c^{\gamma}$. The derivation of this set of vectors with an example is given in \ref{sec:ApKerVecGeneral}. 

Despite this extension, calculating the kernel of $L_c^{\gamma}$ for non-Cartesian meshes involves performing a singular value decomposition, for which the computational cost grows exponentially with the number of grid points as $\mathcal{O}(N_{grid}^3)$, which quickly becomes unaffordable \cite{Golub1996}. Moreover, for most meshes the rank of the kernel reduces to one if the parity and the orthogonal faces disappear, which is nearly always the case for any unstructured mesh, leaving only the constant mode vector. Therefore using the kernel of $L_c^{\gamma}$ to define and quantify the checkerboard problem is often insufficient, leading to the search for a more generally applicable definition.

\subsection{Applying a general definition of checkerboarding}
In the previous section a very restrictive definition of checkerboarding was used, for which some easy examples were given in which case the definition would not be fruitful. A more useful definition can be found if the problem is regarded at control volume level. The essential problem of the decoupled control volume is that the pressure gradient over a given face, $[G\pvec_c]_f$, might give a significant non-zero value, while the value at the adjacent cell $[G_c\pvec_c]_i$ can be (close to) zero. This problem might occur only in a few cells and therefore lie mostly outside of the kernel of $L_c$, that is, if the kernel even contains spurious vectors. 

It turns out that the ratio between the norms of vectors $G_c\pvec_c$ and $G\pvec_c$ gives a good global indication of this decoupling. The expression for the norms of cell-centered and face-centered fields is given by $\rVert a_c \lVert = a_c^T\Omega_ca_c$ and $\rVert a_s \lVert = a_s^T\Omega_sa_s$, respectively. Furthermore, in the lower limit the field lies fully inside the kernel, resulting in a ratio of zero, whereas a perfectly smooth field gives the upper limit, resulting in a ratio of one. Since a higher coefficient should indicate more prevalent checkerboarding, the checkerboard coefficient is finally found by subtracting this ratio from one, resulting in:

\begin{equation}
\label{eq:Ccb}
\begin{split}
    C_{cb}\br{\pvec_c} &= 1 - \frac{\lVert G_c\pvec_c \rVert}{\lVert G\pvec_c \rVert}, \\
    &= 1 - \frac{\pvec_c^TG_c^T\Omega G_c\pvec_c}{\pvec_c^TG^T\Omega_sG\pvec_c}, \\
    &= \frac{\pvec_c^T\br{L-L_c}\pvec_c}{\pvec_c^TL\pvec_c}.
\end{split}
\end{equation}

\noindent $C_{cb}\br{\pvec_c}$ is defined as zero for the constant pressure field in which case $\lVert G\pvec_c \rVert = 0$. The checkerboard coefficient can be calculated for any collocated scalar field. As pressure field oscillations are the focus of this work, $C_{cb}\br{\pvec_c}$ will henceforth simply be denoted by $C_{cb}$. Aside from being a non-dimensional and normalised coefficient, rewriting the coefficient as done in equation \eqref{eq:Ccb} reveals some other properties which support this definition. Firstly, it reveals that the magnitude of the pressure fields does not matter, since the constant pressure vector lies inside the kernel of $L$ and $L_c$ and will therefore be filtered in the operation of equation \eqref{eq:Ccb}. This should always be the case for any definition that is employed, since the value of pressure is not what matters in incompressible flows, rather its gradient. 

Secondly, the matrix $\br{L-L_c} = M\br{I - \Gamma_{cs}\Gamma_{sc}}G$ is innately linked to the fundamental problem of the collocated grid arrangement, which is that the gradient of pressure and the velocities are not defined at the same location. The ensuing interpolators will always apply some smoothing to a field such that only $I\approx\Gamma_{cs}\Gamma_{sc}$. 

Thirdly, this definition has a physical meaning, since the numerical errors in the convective and pressure terms, as discussed in sections \ref{sec:numdiss_wsWIM} and \ref{sec:numdiss_csWIM} respectively, are proportional to the term $\pm\br{L-L_c}\pvec_c$. For the wide-stencil WIM, $\br{L_c-L}\pvec = M\uvec_s$, which is non-zero and leads to the error expressed in equation \eqref{eq:convErr}. Similarly, for the compact-stencil WIM, $\br{L_c-L}\pvec = M_c\uvec_c$, which is non-zero and leads to the pressure error as seen in equation \eqref{eq:presErr}. One notable difference is that $C_{cb}$ is independent of the time-step, which is also a necessary property, since it should be possible to calculate the coefficient without any knowledge of the temporal discretisation.

Finally, $C_{cb}$ is able to detect local oscillations which lie outside of the kernel of $L_c$, which it should be able to do, as mentioned before. A very simple illustration of this argument would be a one-dimensional periodic domain with 6 uniform nodes, containing a specific oscillation. The following calculations would then apply:

\newsavebox\labelbox
\savebox\labelbox{$\begin{matrix}
\refstepcounter{equation}(\theequation)\label{aa}\\
\refstepcounter{equation}(\theequation)\label{bb}\\
\refstepcounter{equation}(\theequation)\label{cc}\\
\refstepcounter{equation}(\theequation)\label{dd}
\end{matrix}$}

\[
    \begin{matrix*}[r]
        \pvec_c = & [0 & 0 & 1 & 0 & \shortminus1 & 0], \\
        \pvec_{c(1,\gamma)}^- = & [1 & \shortminus1 & 1 & \shortminus1 & 1 & \shortminus1], \\
        G\pvec_c = & [0 & 1 & \shortminus1 & \shortminus1 & 1 & 0], \\
        G_c\pvec_c = & [0 & \frac{1}{2} & 0 & \shortminus1 & 0 & \frac{1}{2}],
    \end{matrix*}
\eqno
\usebox{\labelbox}    
\]

\begin{align}
    \pvec_c^T \pvec_{c(1,v)}^- &= 0, \\
    C_{cb} &= 1 - \frac{\lVert G_c\pvec_c \rVert}{\lVert G\pvec_c \rVert} = \frac{5}{8},
\end{align}

\noindent where column vectors are represented horizontally for readability. Intuitively, this wiggle should not have a zero value when quantifying the checkerboard problem, therefore $C_{cb}$ gives a more desirable outcome than the kernel vector method.

\subsection{One possible application of the checkerboard coefficient}
Since $C_{cb}$ gives a global, non-dimensional, normalised coefficient for checkerboarding, it can directly be applied in the solver algorithm to diminish the occurrence of the problem. It could, for example, be used to shift between the possible fractional step method algorithms displayed in table \ref{tab:FSM}, where choices are made concerning: (i) the width of the Laplacian stencil, (ii) the interpolation method for the velocity correction and (iii) the calculation of the pressure predictor. Since most collocated finite volume fractional step method codes apply the compact-stencil WIM, and since the inclusion of the pressure predictor is a known cause of checkerboarding, one way to use $C_{cb}$ is to deliver a negative feedback through this predictor value. To this end, the following expression for $\pvec_c^p$ is used:

\begin{align}
    \pvec_c^p &= \theta_p\pvec_c^n, \label{eq:ppred} \\
    \theta_p &= 1 - C_{cb} = \frac{\pvec_c^{n^T}L_c\pvec_c^n}{\pvec_c^{n^T}L\pvec_c^n}. \label{eq:thetady}
\end{align}

\noindent By doing so, the solver will converge to $\theta_p = 1$ in absence of checkerboarding in the pressure field, benefiting from the lower numerical dissipation this offers. Whereas if the case or mesh is prone to checkerboarding, the algorithm will tend to $\theta_p = 0$, in which case the increase in numerical dissipation can damp the oscillations. This feedback onto the pressure predictor offers a dynamical balance between two problems, so that the user of the code does not have to decide which problem will be more prevalent, creating a unified solver that should perform well in both situations. For this initial attempt at establishing such a negative feedback, a simple linear relation is used to explore its effects, other relations might also work and could give better results.

\newcommand{\colw}{0.46}
\newcommand{\figh}{0.78}

\section{Results}
\label{sec:results}
\FloatBarrier

In this section, the new solver with the predictor pressure given by equations \eqref{eq:ppred} and \eqref{eq:thetady} is tested and compared to the Chorin and Van Kan methods, resulting in the three solvers given by table \ref{tab:solvers}. To do so, firstly, numerical dissipation is measured using a two-dimensional Taylor-Green vortex. Secondly, a temporal and spatial convergence study is performed using a two-dimensional lid-driven cavity. And finally, a turbulent channel flow case is used to measure overall accuracy of the solver and to study the behaviour of the checkerboard coefficient in unsteady flows. All cases were run with OpenFOAM, using the symmetry-preserving spatial discretisation and Runge-Kutta 3 temporal integration, which are both implemented in the \emph{RKSymFoam} solver developed for \cite{Komen2021, Hopman2023b}.

\begin{table}[ht]
\centering
\caption{Settings of the tested solvers}
\begin{tabular}{cccc}
\label{tab:solvers}
& $\theta_0$ & $\theta_1$ & $\theta_{dy}$ \\
\midrule
$\theta_p$ & $0$ & $1$ & $1-C_{cb}$ \\
\end{tabular}
\end{table}

In each case the newly implemented solvers are compared to a readily implemented solver available in OpenFOAM, which is either \emph{icoFoam} for laminar flows or \emph{pisoFoam} for turbulent flows \cite{greenshields2023}. Apart from some choices in spatial discretisation, the biggest difference of the OpenFOAM solvers is how they remove time-step dependency of the compact-stencil WIM, which is similar to and derived from the methods used by \cite{Choi1999, Yu2002}. For explicit time-stepping it is given by:

\begin{align}
    \uvec_s^p &= \Gamma_{cs}^\gamma\uvec_c^p - C_{u_s}\uvec_{s,corr}, \\
    \uvec_{s,corr} &= \uvec_s^n - \Gamma_{cs}^\gamma\uvec_c^n, \\
    [C_{u_s}]_{f,f} &= 1 - min\left(\frac{||[\uvec_{s,corr}]_f||}{||[\uvec_s^n]_f|| - \epsilon}, 1\right),
\end{align}

\noindent where $\epsilon$ is a very small number to avoid division by zero. This term stabilises the results at the cost of introducing a sizable amount of numerical dissipation, which was shown by \cite{Vuorinen2014}.

\FloatBarrier

\subsection{Numerical dissipation}
\label{sec:res_TGV}
\begin{figure}[ht]
    \centering
    \includegraphics[trim={0 0 7.15cm 0},clip, width=0.5\textwidth]{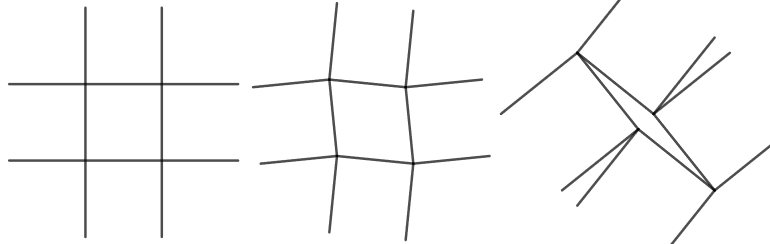}
    \caption{Graphical representation of uniform (left) and perturbed (right) meshes.}
    \label{fig:TGV_mesh}
\end{figure}

A two-dimensional Taylor-Green vortex case was used to examine the balance between numerical dissipation and checkerboarding of each solver. The solutions to the pressure and velocity fields can be calculated analytically, in non-dimensional units, as \cite{Taylor1937}:

\begin{align}
    u_x(x,y,t) &= \sin(x)\cos(y)e^{\shortminus2\nu t}, \label{eq:TGV_ux} \\
    u_y(x,y,t) &= \shortminus\cos(x)\sin(y)e^{\shortminus2\nu t}, \label{eq:TGV_uy} \\
    p(x,y,t)   &= \frac{1}{4}(\cos(2x)+\cos(2y))e^{\shortminus4\nu t} + C,
\end{align}

\noindent where $\mathbf{u}$, $L$, $\nu$, $t$ and $p$ are non-dimensionalised using: Maximum velocity, $U_c=1$, characteristic length, $L_c = \frac{L_D}{2k\pi}$ with domain length $L_D$ and wave number $k$, $\nu_c=U_cL_c$, $t_c=L_c/U_c$ and $p_c = U_c^2/\rho$, respectively. $C$ can be any constant since absolute pressure has no meaning in an incompressible case with these boundary conditions, and is set to $0$. The case was run on a square domain with sides of length $L_D$ with periodic boundaries and using $k = 1$. The domain is discretised in $x-$ and $y-$directions with $N_c = 33$ control volumes, so that the pressure at the central control volume can be used as a reference cell with exactly $p_{Ref} = 0.5$. The time-step size was chosen as $\Delta t = 0.05$ to meet the Courant–Friedrichs–Lewy (CFL) condition, and the total simulated time was 10 time units. The case was run with zero viscosity and with $Re = 100$, based on $U_c$ and $L_c$. Furthermore, a uniform and a perturbed mesh were used, of which the latter has slightly skewed control volumes to provoke checkerboarding, see figure \ref{fig:TGV_mesh}. 

The results are quantified in terms of checkerboarding using $C_{cb}$, as given by equation \eqref{eq:Ccb}. In terms of numerical dissipation, the pressure diffusion is calculated and taken as a measure of accuracy. It is almost identical to the pressure error term, which is the only term that can have non-physical contributions to the total kinetic energy when using the compact-stencil WIM, as explained in section \ref{sec:numdiss_csWIM}. The pressure diffusion is normalised by the characteristic time, $t_c$, and the analytical value for the total kinetic energy, $E_{K,Ana}$, which can be found by integrating equations \eqref{eq:TGV_ux} and \eqref{eq:TGV_uy} over the full domain:

\begin{align}
    E_{K,Ana} &= E_0e^{\shortminus4\nu t}, \\
    \varepsilon_{pDif} &= \frac{\mathbf{u}_c^T\Omega G_c\mathbf{p}_c}{E_{K,Ana}/t_c}.
\end{align}

\begin{figure}[ht]
    \centering
    \includegraphics[width=\textwidth]{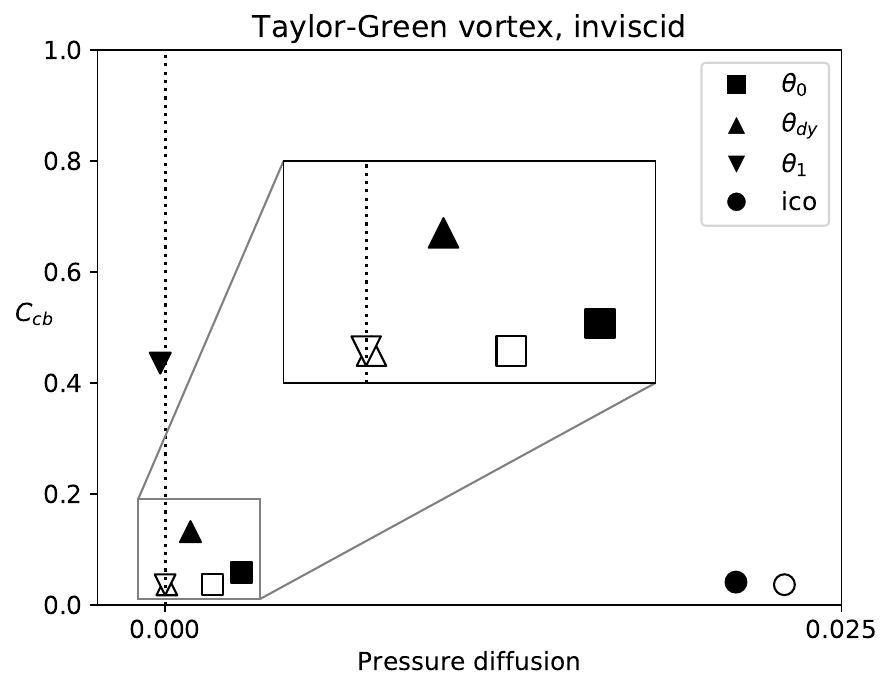}
    \caption{Balance between checkerboarding and pressure diffusion in an inviscid Taylor-Green vortex on uniform (unfilled markers) and perturbed (filled markers) meshes.}
    \label{fig:TGV_invisc}
\end{figure}

For the inviscid case, the results are plotted in figure \ref{fig:TGV_invisc}. As the global pressure diffusion should be exactly zero, any methods to the right of the dotted line introduce numerical dissipation, whereas values to the left can become unstable. The results for the case on a uniform mesh show very little checkerboarding. The $\theta_1$ method introduces practically no numerical dissipation since the pressure error is greatly reduced. The $\theta_0$ method does show this error, although much smaller than the one introduced by \emph{icoFoam}. The $\theta_{dy}$ method gives practically the same result as the $\theta_1$ method, since there is almost no checkerboarding and $\theta_p$ has converged close to $1$.
When skewness is introduced to the mesh to provoke checkerboarding, the trade-off between numerical dissipation and checkerboarding becomes clear. The $\theta_1$ method has moved to the unstable side because the pressure error is introducing energy, and checkerboarding is clearly present. This may be related to the fact that the perturbed volumes are not well-centered, a property which is important to guarantee stability \cite{Santos2022,Santos2023}. Even though the method is on the unstable side because it is adding kinetic energy, in practice the simulation is still running stably. On the contrary, \emph{icoFoam} is still the most dissipative method, however, it has maintained low levels of checkerboarding. The $\theta_0$ method lies in between the former methods, while the $\theta_{dy}$ method again lies between the $\theta_1$ and $\theta_0$ methods, as $C_{cb}$ has settled around $0.15$ and is balancing the trade-off between checkerboarding and numerical dissipation.

\begin{figure}[ht]
    \centering
    \includegraphics[width=\textwidth]{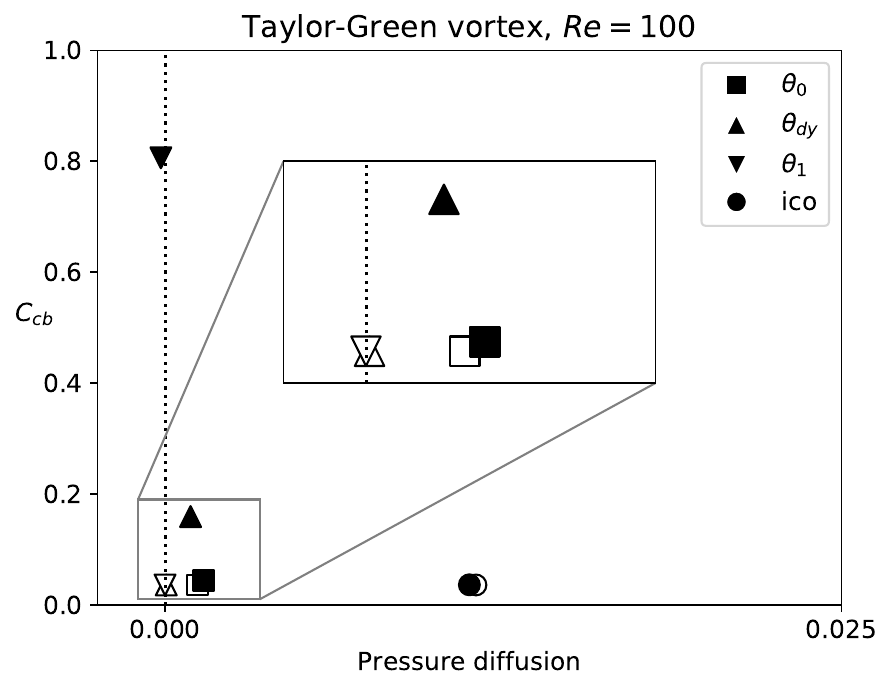}
    \caption{Balance between checkerboarding and pressure diffusion in Taylor-Green vortex at $Re = 100$ on uniform (unfilled markers) and perturbed (filled markers) meshes.}
    \label{fig:TGV_visc}
\end{figure}

When viscosity is introduced to the case, the results slightly change, as show in figure \ref{fig:TGV_visc}. Again, the uniform mesh shows practically no checkerboarding and the results are similar to those of figure \ref{fig:TGV_invisc}. Skewing the mesh introduces checkerboarding like before. The $\theta_1$ method is still on the unstable side, which may have caused an even greater occurrence of checkerboarding. The $\theta_0$ method again nicely shows the balance between numerical dissipation and checkerboarding, as \emph{icoFoam} shows less checkerboarding at the price of more numerical dissipation. Finally, the $\theta_{dy}$ method again has settled around $C_{cb} = 0.15$, in between the $\theta_1$ and $\theta_0$ methods where it facilitates a trade-off between checkerboarding and numerical dissipation. 

\begin{figure}[ht]
    \centering
    \includegraphics[trim={6cm 23cm 4cm 20cm}, clip, width=\textwidth]{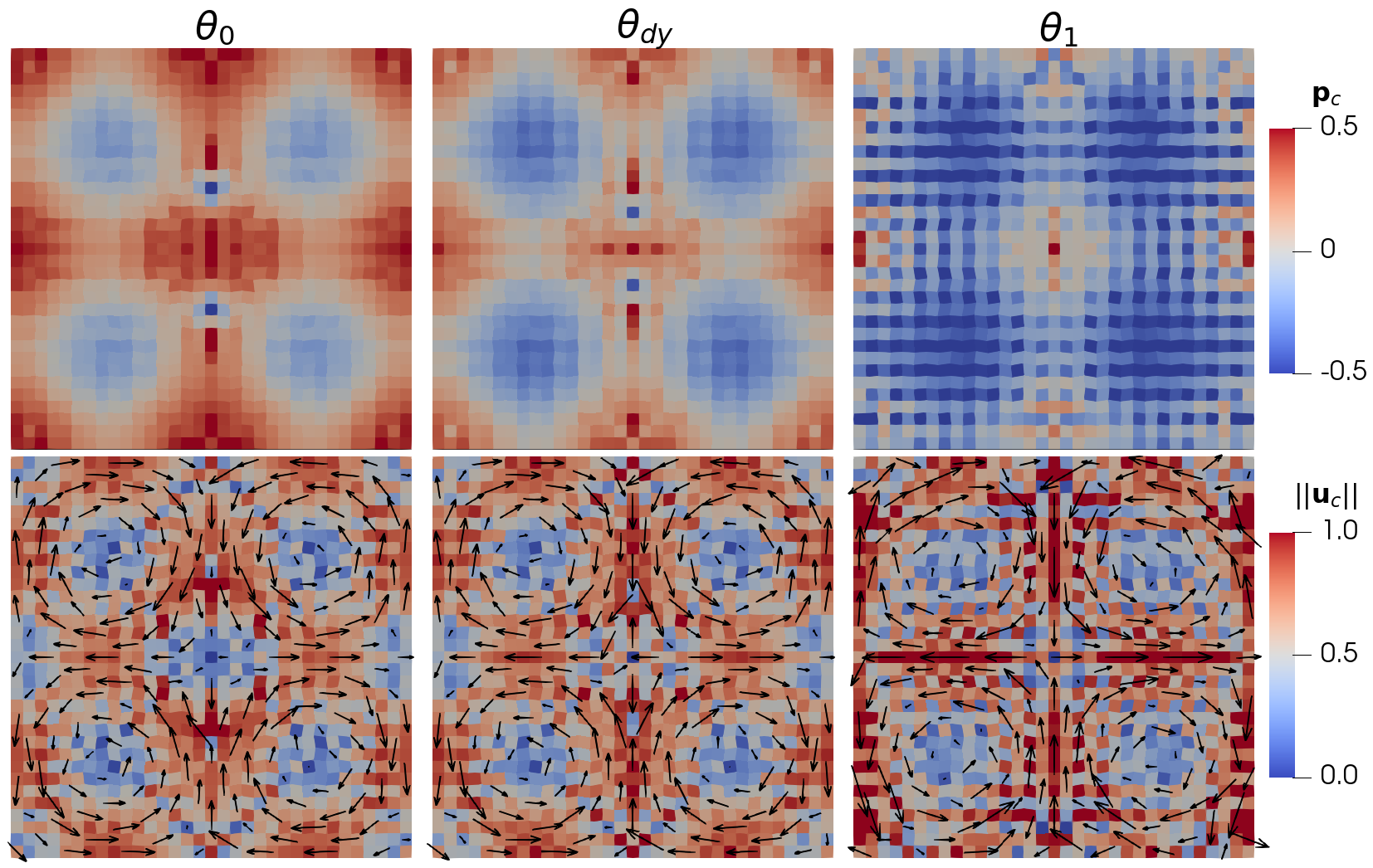}
    \caption{Qualitative comparison of checkerboard intensity for the inviscid Taylor-Green vortex on a perturbed mesh.}
    \label{fig:TGV_qualitative}
\end{figure}

Finally, the effective balancing of the amount of checkerboarding can be seen qualitatively in figure \ref{fig:TGV_qualitative}. In this figure, the effect of the pressure predictor on the pressure and velocity fields in the inviscid case on a perturbed mesh can be seen. Oscillations are present in each case, with the biggest difference between the $\theta_0$ and $\theta_1$ methods, as also quantified by $C_{cb}$. $\theta_{dy}$ is able to dynamically settle at a value between these methods, where oscillations are slightly higher with respect to the $\theta_0$ method, but not yet as problematic as in the $\theta_1$ method.

\FloatBarrier

\subsection{Convergence studies}
\FloatBarrier

The pressure error for the $\theta_0$ method is of order $\mathcal{O}\br{\Delta t\Delta x^2}$, whereas for the $\theta_1$ method it has an order of $\mathcal{O}\br{\Delta t^2\Delta x^2}$ \cite{Chorin1967}. Since this property can greatly reduce the computational cost and improve accuracy of the solution, it is important to study the effects of the dynamical pressure predictor on this order of accuracy.
To this end, a lid-driven cavity flow case was used to study the effects of spatial and temporal discretisation on the numerical error and checkerboarding. The case consists of a square domain with sides $L_D$, of which the top is moving with velocity $U_L$. No-slip boundary conditions were used for the velocity field, while a zero-gradient pressure boundary condition was used. Variables were non-dimensionalised in the same way as in section \ref{sec:res_TGV}, with the exception of $L_c = L_D$. The case was run at $Re = 1000$. Only uniform meshes were used, as checkerboarding can be observed without mesh perturbation, which might be related to the difference in boundary condition, as it is the main difference between both cases. Two different mesh refinements were tested, a coarse mesh of $64\times64$ control volumes and a fine mesh of $512\times512$ control volumes. This was done to see if the effect of the pressure error diminished with mesh refinement, as is to be expected. The case was run with time-step sizes of $\Delta t = 2^k\delta t$ up to a total simulated time of $2^6\delta t$, with $k = 0, 1,\dots, 6$ and $\delta t = 4\cdot10^{-6}$. These time-step sizes were chosen such that the largest time-step still meets the CFL condition on the fine mesh. The convergence of numerical dissipation was measured by measuring the difference in kinetic energy, $\varepsilon_{E_k}$,between the solution of the particular time-step size, compared to the solution of the same solver on the same mesh with the smallest time-step, $\delta t$.

For this section, an additional solver $\theta_{\frac{1}{2}}$ was introduced with $\theta_p = \frac{1}{2}$. This was done because $\theta_{dy}$ has a dependence on grid size and time-step, and will therefore not be constant throughout the convergence studies. While studying the results for $\theta_{dy}$ it was found that for these cases $\frac{1}{2}$ is a typical value for $C_{cb}$, which was therefore chosen to grants insight into a constant intermediate value of $\theta_p$.

\subsubsection*{Convergence of the numerical error}
\begin{figure}[ht]
    \centering
    \includegraphics[width=0.75\textwidth]{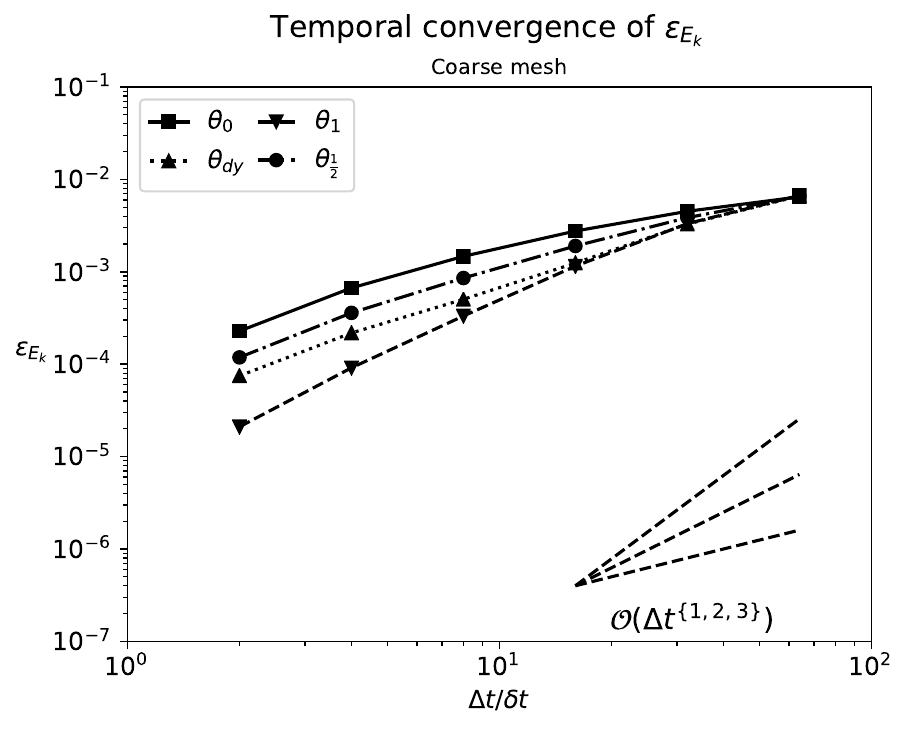}
    \caption{Temporal convergence of the numerical error in kinetic energy for the different implemented solvers, on a coarse $64 \times 64$ mesh.}
    \label{fig:LDC_temp_EK_64}
\end{figure}

The results for the coarse mesh are shown in figure \ref{fig:LDC_temp_EK_64}. The pressure error should become more dominant as the mesh becomes coarser. The pressure error is visible as the $\theta_0$ method shows a first-order relation of the numerical error with respect to $\Delta t$, whereas the $\theta_1$ method shows a second-order relation, as expected when the pressure error is dominant. The methods with intermediate values for $\theta_p$ show first-order behaviour, with the $\theta_{\frac{1}{2}}$ and $\theta_{dy}$ methods showing some conservation of the second-order relation. This can be explained by the fact that the second-order relation can only be achieved if $\pvec_c'$ approximates $\Delta t \frac{d\pvec_c}{dt} = \pvec_c^{n+1} - \pvec_c^p$, which is the case only when $\theta_p = 1$ and $\pvec_c^p = \pvec_c^n$. In practice this could mean that $\theta_1$ is favourable on coarse meshes unless a certain threshold of checkerboarding is surpassed, so that the second-order relation between the pressure error and time-step size can be maintained, instead of a gradual decay in $\theta_p$. 

\begin{figure}[ht]
    \centering
    \includegraphics[width=0.75\textwidth]{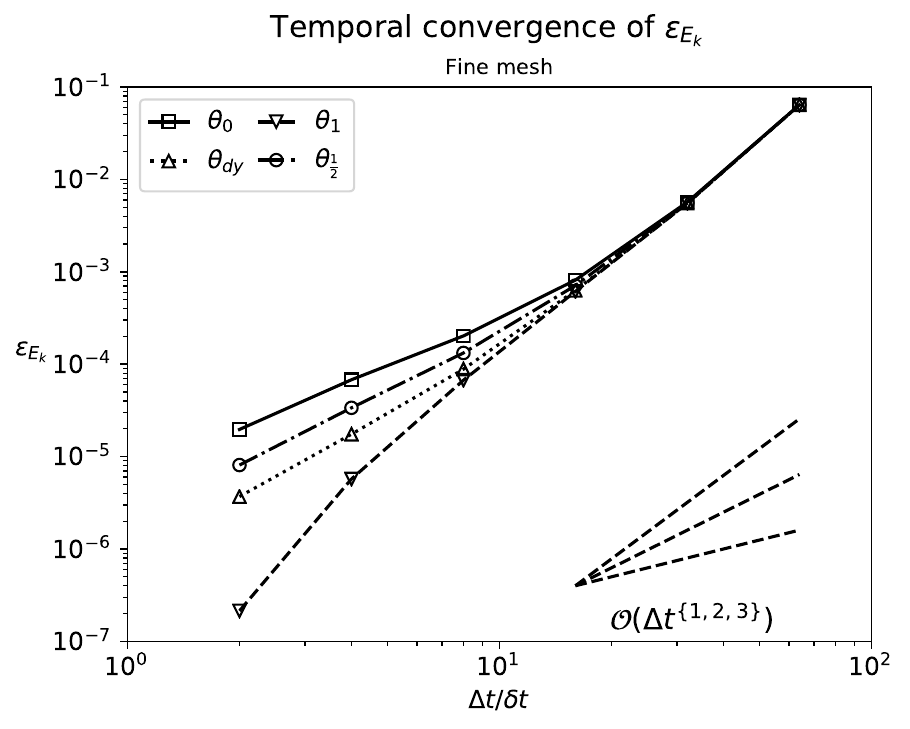}
    \caption{Temporal convergence of the numerical error in kinetic energy for the different implemented solvers, on a coarse $512 \times 512$ mesh.}
    \label{fig:LDC_temp_EK_512}
\end{figure}

On a finer mesh, the pressure error becomes less dominant, which can be seen from figure \ref{fig:LDC_temp_EK_512}. As the pressure error is less dominant, the temporal error dominates the order of the relation between the numerical error and the time-step size. Since the temporal integration method used was the classical Runge-Kutta 3 method, the third-order relation is retrieved. The $\theta_1$ method shows this relation most clearly, whereas the other methods seem to have a slightly lower order relation, with the $\theta_{dy}$ and $\theta_{\frac{1}{2}}$ methods showing intermediate values again. This can be caused by the fact that the pressure error still has a small contribution to the overall relation between time-step size and numerical error, and the full order of the temporal error is not yet showing on this mesh.

\subsubsection*{Convergence of the checkerboard coefficient}
\begin{figure}[ht]
    \centering
    \includegraphics[width=0.75\textwidth]{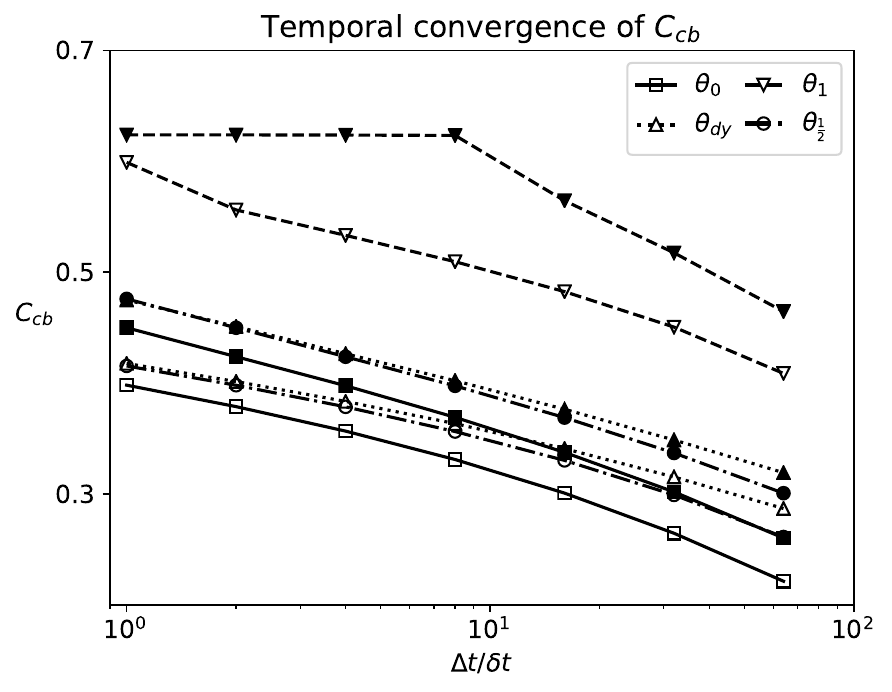}
    \caption{Temporal convergence of checkerboarding in kinetic energy for the different implemented solvers on a coarse (filled markers) and a fine (unfilled markers) mesh.}
    \label{fig:LDC_temp_pcb}
\end{figure}

As was noted in section \ref{sec:occurrence} a very small time-step size can be the cause of checkerboarding. Moreover, a fine mesh will be able to capture higher frequency modes in the pressure field, which can then be resolved by both $L$ and $L_c$ and therefore not contribute to the overall value of $C_{cb}$. The influence of the mesh and time-step size on $C_{cb}$ can be seen in figure \ref{fig:LDC_temp_pcb}. First of all, it can be noted that checkerboarding is significantly higher for the $\theta_1$ method and gets lower for the $\theta_{\frac{1}{2}}$ method down to the lowest values for the $\theta_0$ method. A smaller time-step does indeed contribute to the prevalence of checkerboarding, although the value does seem to culminate around $0.6$ for the $\theta_1$ method, which might be specific to this case or mesh. Moreover, checkerboarding is more prevalent on coarse meshes, confirming the expectations. Finally, the $\theta_{dy}$ method shows a decay in checkerboarding upon increasing the time-step size, which is not as steep as for the other methods. This can be explained by the fact that lower checkerboarding has a negative feedback on the pressure predictor, increasing $\theta_p$, therefore slowly moving towards the values of the $\theta_1$ method. In conclusion, the influence of time-step size and grid size are confirmed, while the dynamic and balancing behaviour of the $\theta_{dy}$ method is visualised. 

\FloatBarrier

\subsection{Channel flow}
\FloatBarrier

A turbulent channel flow case at $Re_{\tau} = \frac{u_{\tau}h}{\nu} = 180$ was used to further examine the occurrence and behaviour of checkerboarding in a transient case. The origins of checkerboarding by $\Delta t\rightarrow 0^+$ and $\ptil_c^p = \ptil_c^n$ both rely on minimal changes in $\ptil_c'$ and effectively solving the Poisson equation for the same solution iteratively, which should be more pronounced in laminar, steady cases. Therefore, it is interesting to see what happens when the case becomes unsteady. 
The quantified checkerboarding will be compared between the suggested solvers, and the effect of mesh refinement will be studied. Furthermore, the accuracy of the solvers will be measured and compared using turbulent kinetic energy budgets and the influence of the temporal scheme and mesh will be taken into consideration.

The case was run inside a domain of dimensions $[4\pi h \times 2h \times \frac{4}{3}\pi h]$ in $x,y,z$-directions, respectively, with channel half-height $h$, which are the same dimensions as the ones used for the domains in \cite{Vreman2014, Komen2020}. Several Cartesian grids where used with uniform spacing in $x,z$-directions and stretching in $y$-direction so that the vertices are placed according to equation \eqref{eq:chanGridStretch}, as in \cite{Zhang2015}, with $g_y = 1.85$. The specifics of these meshes can be seen in table \ref{tab:CF_mesh}. Non-dimensionalisation is similar to \ref{sec:res_TGV}, with $L_c = h$, the channel half-height, and $U_c = u_{\tau}$, the frictional velocity, given by $u_{\tau} = \sqrt{\frac{\tau_w}{\rho}}$, in terms of wall shear stress, $\tau_w = -h\nabla p$, or $u_{\tau} = \sqrt{\nu\left(\partial_y\langle u\rangle\right)_w}$, in terms of wall-derivative of the mean stream-wise velocity at the wall. The flow was developed for 20 time units and continued for up to 80 time units to gather the results.

\begin{equation}
\label{eq:chanGridStretch}
    y_j = \frac{L_y}{2} \left(1 + \frac{tanh\left(g_y \frac{2j - N_y}{N_y}\right)}{tanh\left(g_y\right)}\right), \quad j=0,1,\dots,N_y
\end{equation}

\begin{table}[ht]
\centering  
\begin{tabular}{lcccccccc}
\toprule
& {$N_x$} & {$N_y$} & {$N_z$} & {$N_{tot}$} & {$\Delta x^+$} & {$\Delta y^+_B$} & {$\Delta y^+_w$} & {$\Delta z^+$} \\
\midrule
$X_{40}$  & 40  & 80  & 40  & 128,000    & 56.549 & 8.741 & 0.861 & 18.850 \\
$X_{80}$  & 80  & 100 & 80  & 640,000    & 28.274 & 6.994 & 0.683 & 9.425  \\
$X_{120}$ & 120 & 120 & 120 & 1,728,000  & 18.850 & 5.830 & 0.566 & 6.283  \\
$X_{180}$ & 180 & 120 & 120 & 2,592 000  & 12.566 & 5.830 & 0.566 & 6.283  \\
DNS\cite{Vreman2014}    & 384 & 193 & 192 & 14,229,504 & 5.89   & 2.945 & 0.024 & 3.93   \\
\bottomrule
\end{tabular}
\caption{Different mesh characteristics used for the channel flow case.}
\label{tab:CF_mesh}
\end{table}

\subsubsection*{Quantified checkerboarding}
The checkerboarding was quantified and monitored for each solver and compared to the readily available OpenFOAM solver \emph{pisoFoam}, which is used for transient incompressible flows. The quantification through $C_{cb}$ can be seen in table \ref{tab:CF_pcb}. The same conclusions with respect to mesh refining can be taken from this case as before, i.e. mesh refining decreases the amount of checkerboarding. Between the $X_{120}$ and $X_{180}$ meshes, the checkerboarding does not seem to decrease much further, which indicates that the level of checkerboarding does not converge to $0$ with mesh refinement and some checkerboarding is due to the solver itself. The amount of checkerboarding is highest for the $\theta_1$ method and is decreased greatly by the $\theta_0$ method, especially on coarse grids. As expected and showed before, the $\theta_{dy}$ method dynamically balances between these values through its negative feedback on the pressure predictor. 

\begin{table}[ht]
\centering  
\begin{tabular}{lcccc}
\toprule
& $X_{40}$ & $X_{80}$ & $X_{120}$ & $X_{180}$\\
\midrule
$\theta_0$    & 0.57 $\pm$ 0.014 & 0.30 $\pm$ 0.009 & 0.17 $\pm$ 0.004 & 0.16 $\pm$ 0.005 \\
$\theta_{dy}$ & 0.60 $\pm$ 0.013 & 0.34 $\pm$ 0.008 & 0.18 $\pm$ 0.006 & 0.17 $\pm$ 0.006 \\
$\theta_1$    & 0.72 $\pm$ 0.024 & 0.40 $\pm$ 0.013 & 0.20 $\pm$ 0.007 & 0.18 $\pm$ 0.007 \\
piso          & 0.34 $\pm$ 0.025 & 0.15 $\pm$ 0.010 & 0.08 $\pm$ 0.007 & 0.05 $\pm$ 0.030 \\
\bottomrule
\end{tabular}
\caption{Overview of $C_{cb}$ for the different solvers and meshes.}
\label{tab:CF_pcb}
\end{table}

\begin{figure}[ht]
    \centering
    \includegraphics[width=0.8\textwidth]{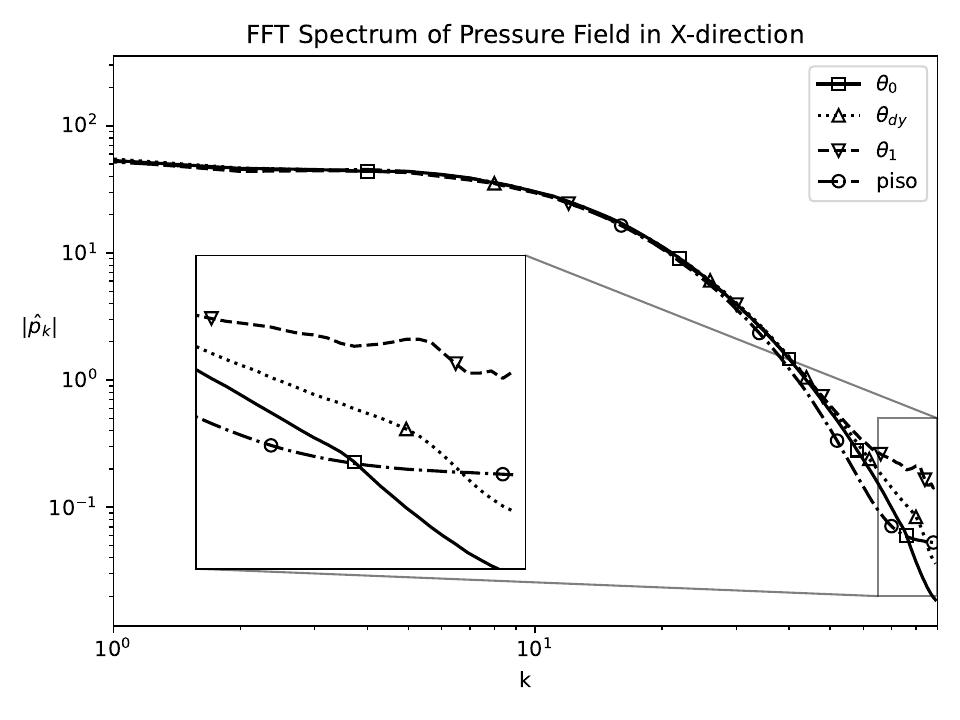}
    \caption{Fast Fourier Transform spectrum on the $X_{180}$ mesh.}
    \label{fig:CF_xspatfft}
\end{figure}

Verification of the quantification method in this case was performed by looking at the Fast Fourier Transform (FFT) spectrum of the pressure field in the stream-wise direction, for the $X_{180}$ mesh. To obtain this spectrum, instantaneous pressure fields of the results with intervals of $0.25$ time units were decomposed into $N_y \times N_z$ individual pressure lines of length $N_x$. The FFT was applied to decompose these lines, after which the resulting magnitudes were averaged and plotted for each wave-number, as seen in figure \ref{fig:CF_xspatfft}. From this graph a clear peak at higher frequencies is visible for the $\theta_1$ method. Compared to the spectra of the other solvers, this peak indicates a non-physical effect, suggesting numerical origins as is the case for checkerboard oscillations. This peak disappears gradually from methods $\theta_{dy}$ to $\theta_0$, which have a smoother trajectory. \emph{pisoFoam} has noticeably lower magnitudes across the upper half of the spectrum, indicating that the solution is damped even in the non-problematic frequencies. The tail of the curve flattens, which also gives a suggestion of numerical interference with the physical solutions. 

\subsubsection{Turbulent kinetic energy budgets}
To assess the quality of the solvers, an analysis was made of the turbulent kinetic energy budgets, along with the first and second-order statistics of the velocity fields. Initially, the comparison was made on the $X_{40}$ mesh using Backward Euler time-stepping, so that comparison to \emph{pisoFoam} was possible. Later, the results are shown for the $X_{180}$ mesh using the Runge-Kutta 3 temporal integration scheme. The solution by \cite{Vreman2014} is used as a reference with a DNS quality mesh. A derivation of the budget terms can be found in \cite{Durbin2011}. 

To find the turbulent kinetic energy terms at cell level, equation \eqref{eq:NSMomDis} is first left-multiplied by $\Omega^{-1}$, then rewritten at a cell level to:

\begin{equation}
    \partial_t u_i = C_{i} + D_{i} + P_{i}.
\end{equation}

\noindent The inner-product at each cell of the fluctuating velocity, $u_i'$, with the fluctuation of this equation leads to the local turbulent kinetic energy budget terms:

\begin{equation}
    \partial_t k = C_{k} + D_{k} + P_{k},
\end{equation}

\noindent with the convective, diffusive and pressure diffusion budget terms on the RHS, respectively. These terms can be exactly discretised and calculated as they are found in the algorithm, without any approximations having to be made, since each term on the RHS is already discretised at cell level. In practice however, the convective and diffusive terms, $C_{k}$ and $D_{k}$ respectively, are often split into their partial budget terms, leading to the transport equation of turbulent kinetic energy:

\begin{equation}
    \frac{Dk}{Dt} = \partial_t k + \lr{u_j}\partial_j k = C_k^P + C_k^T + D_k^v + D_k^{\epsilon} + P_k,
\end{equation}

\noindent with the production, transport, viscous diffusion, dissipation and pressure diffusion budget terms on the RHS, respectively. This split can be done on a continuous level using the chain and product rules for differentiation, which are hard to reflect in a discrete sense without making a choice in discretisation. This can then lead to approximations and a non-zero sum of the budget terms. Instead, in this work, the amount of approximations is minimised by expressing the budget terms as follows:

\begin{equation}
\label{eq:discbuds}
\begin{split}
    C_k &= \lr{u_iC_{i}} - \lr{u_i}\ \lr{C_{i}}, \\
    D_k &= \lr{u_iD_{i}} - \lr{u_i}\ \lr{D_{i}}, \\
    P_k &= \lr{u_iP_{i}} - \lr{u_i}\ \lr{P_{i}}, \\
    C_k^P &= -\big(\lr{u_iu_j} - \lr{u_i}\ \lr{u_j}\big)\lr{\partial_ju_i}, \\
    D_k^{\epsilon} &= -\nu\lr{\partial_ju_i\partial_ju_i} + \nu\lr{\partial_ju_i}\ \lr{\partial_ju_i}, \\
    C_k^T &= C_k - C_k^P + \lr{u_j}\partial_j k, \\
    D_k^v &= D_k - D_k^{\epsilon},
\end{split}    
\end{equation}

\noindent such that the only choice in discretisation that remains to be made, is how to take a cell-centered gradient of the velocity, $\partial_ju_i$. In this work, the cell-centered gradient is taken for each component of $\uvec_c$ separately using $G_c$, creating a tensor field with 9 components at the cell-centers. The derivation of the equations and the implementation of the method in OpenFOAM can be found in the GitHub repository of \cite{Hopman2023c}.

\begin{figure}[ht]
    \centering
    \begin{minipage}[b]{\colw\linewidth}
        \centering
        \includegraphics[height=\figh\linewidth]{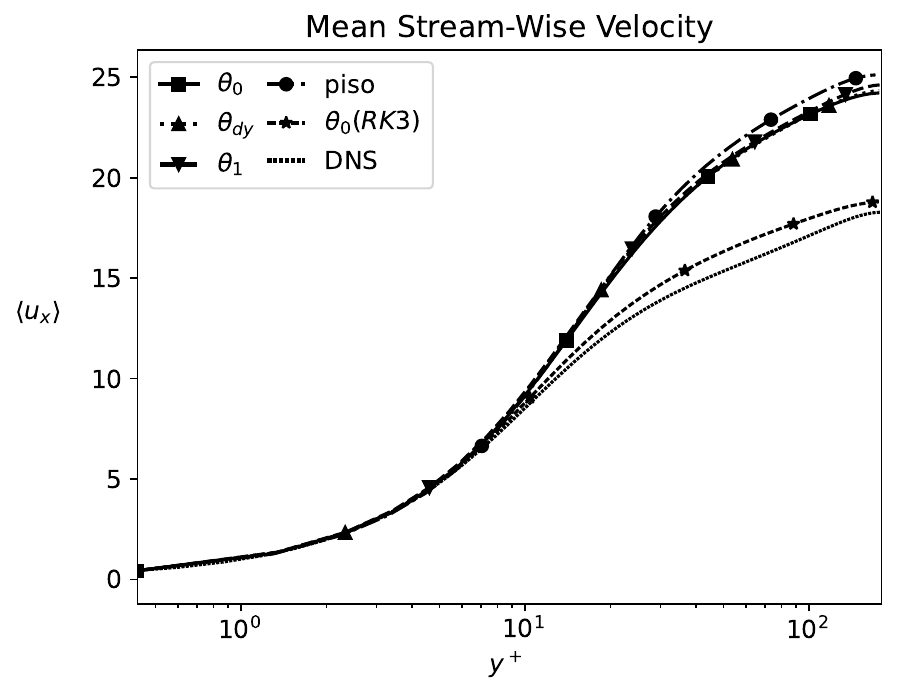}
        \subcaption{}
        \label{fig:CF_Coarse_Ux}
    \end{minipage}
    \hspace{0.05\linewidth}
    \begin{minipage}[b]{\colw\linewidth}
        \centering
        \includegraphics[trim={0 0 0 0.75cm}, clip, height=\figh\linewidth]{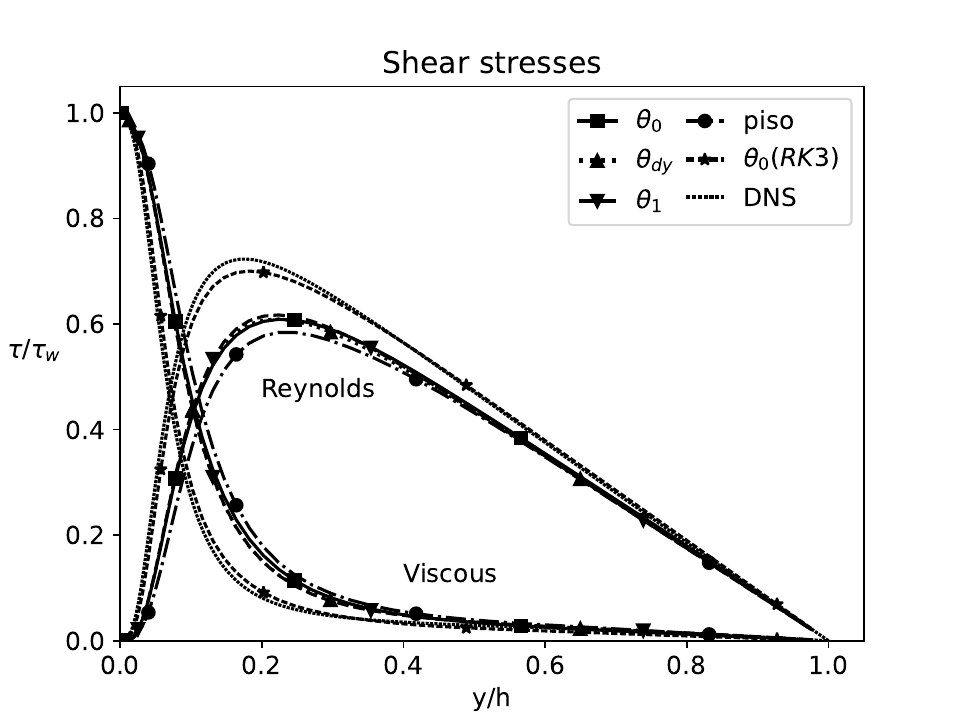}
        \subcaption{}
        \label{fig:CF_Coarse_tau}
    \end{minipage}
    \vspace{0.05\linewidth}
    \begin{minipage}[b]{\colw\linewidth}
        \centering
        \includegraphics[height=\figh\linewidth]{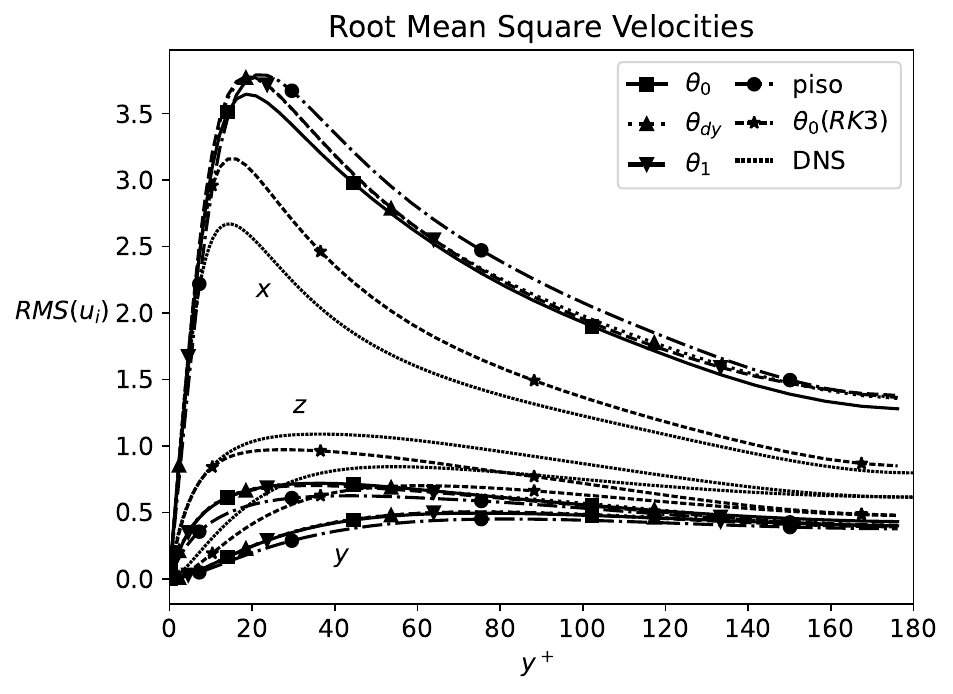}
        \subcaption{}
        \label{fig:CF_Coarse_RMSs}
    \end{minipage}
    \caption{(a) Mean stream-wise velocity on the $X_{40}$ mesh. (b) Profiles of the viscous and Reynolds shear stress on the $X_{40}$ mesh, showing that \emph{pisoFoam} is not able to fully develop turbulence. (c) RMS velocities on the $X_{40}$ mesh. Influence of the temporal integration scheme is shown by plotting the $\theta_0$ method using the RK3 scheme.}
\end{figure}

\begin{figure}[ht]
    \centering
    \begin{minipage}[b]{\colw\linewidth}
        \centering
        \includegraphics[height=\figh\linewidth]{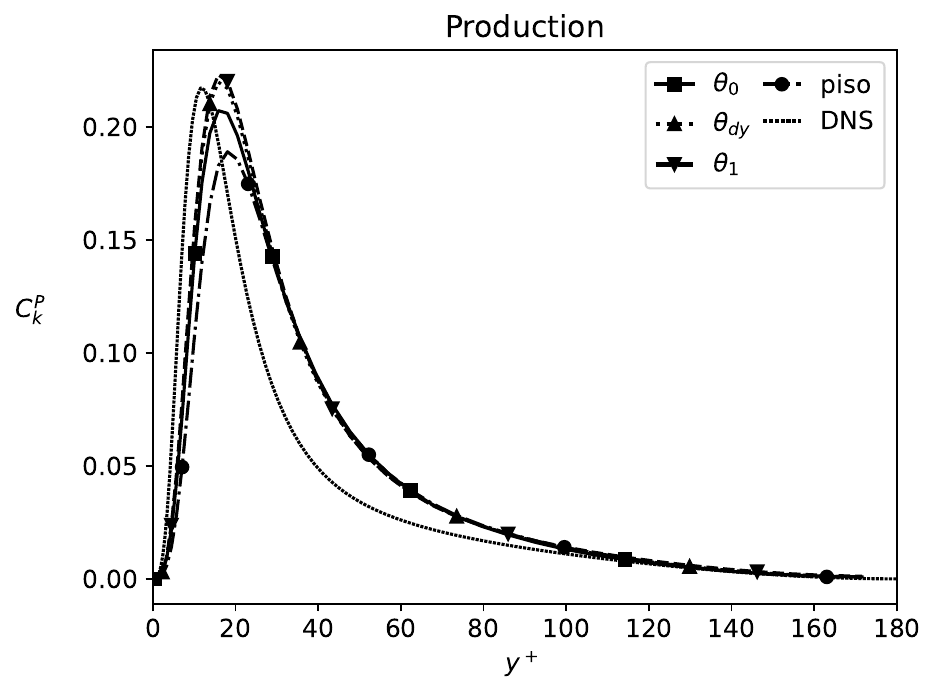}
        \subcaption{}
        \label{fig:CF_Coarse_prod}
    \end{minipage}
    \hspace{0.05\linewidth}
    \begin{minipage}[b]{\colw\linewidth}
        \centering
        \includegraphics[height=\figh\linewidth]{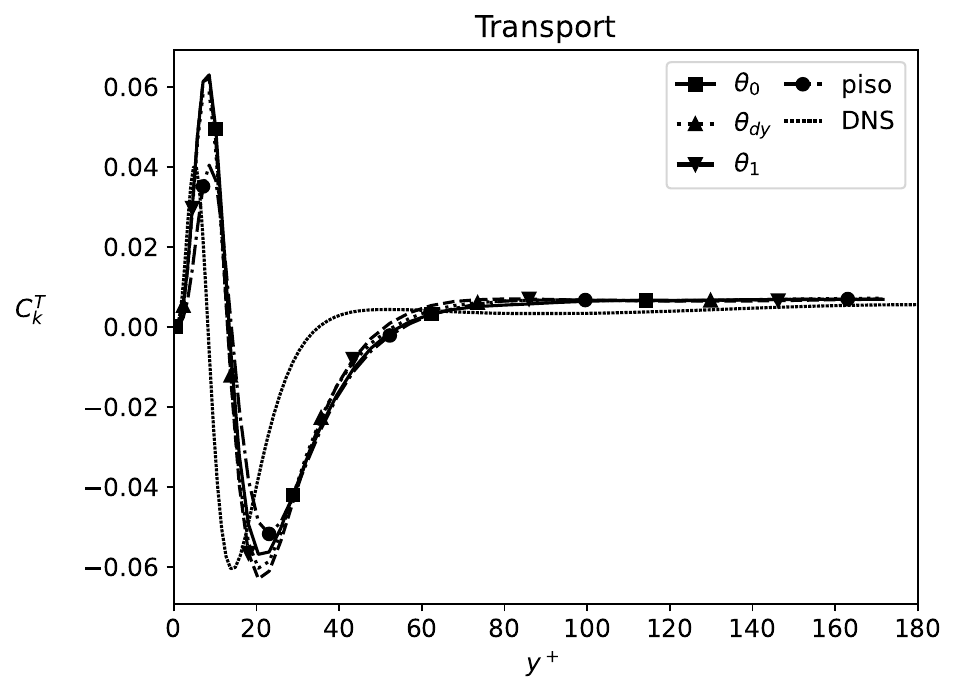}
        \subcaption{}
        \label{fig:CF_Coarse_trans}
    \end{minipage}
    \vspace{0.05\linewidth}
    \begin{minipage}[b]{\colw\linewidth}
        \centering
        \includegraphics[height=\figh\linewidth]{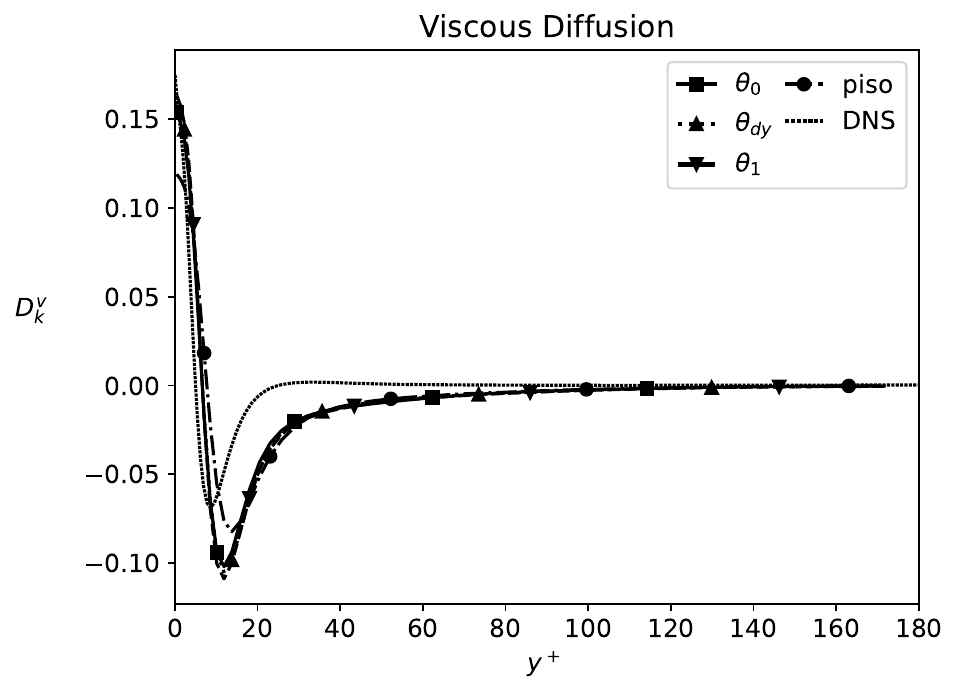}
        \subcaption{}
        \label{fig:CF_Coarse_vDiff}
    \end{minipage}
    \hspace{0.05\linewidth}
    \begin{minipage}[b]{\colw\linewidth}
        \centering
        \includegraphics[height=\figh\linewidth]{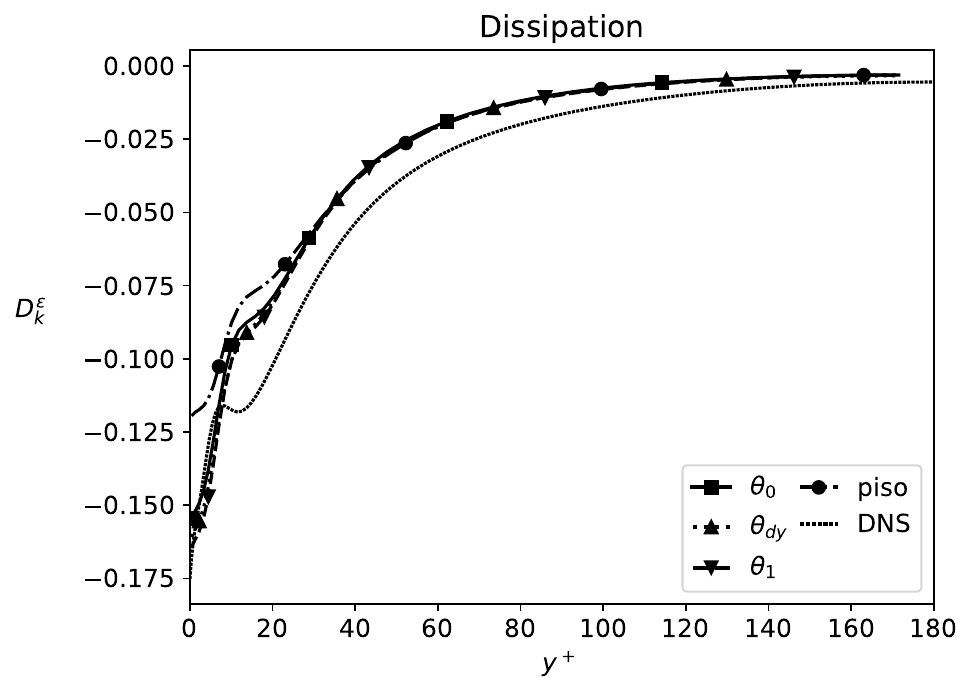}
        \subcaption{}
        \label{fig:CF_Coarse_diss}
    \end{minipage}
    \vspace{0.05\linewidth}
    \begin{minipage}[b]{\colw\linewidth}
        \centering
        \includegraphics[height=\figh\linewidth]{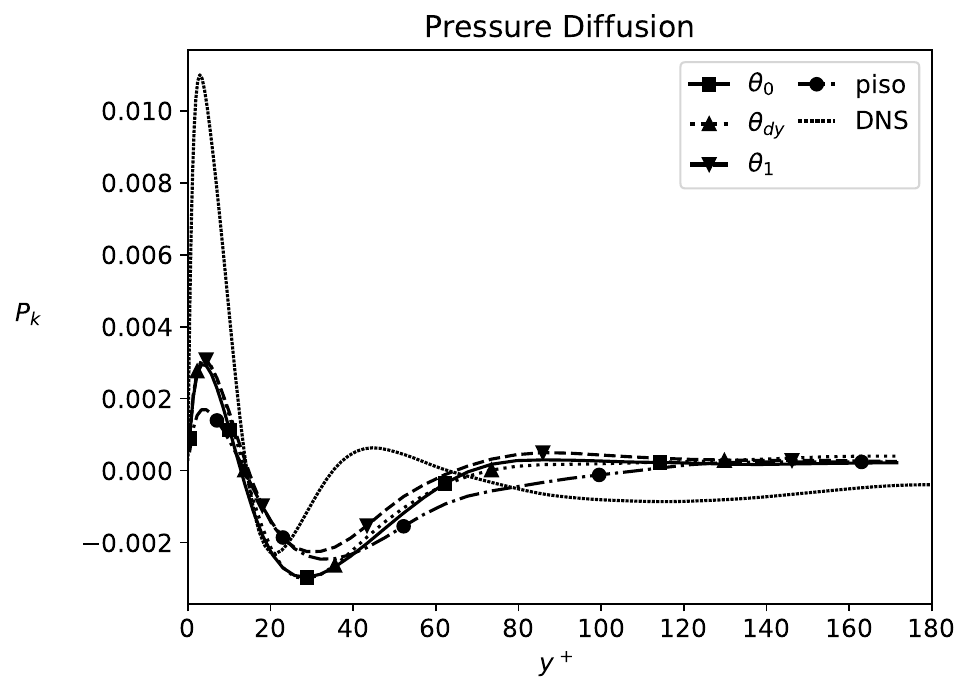}
        \subcaption{}
        \label{fig:CF_Coarse_pDiff}
    \end{minipage}
    \caption{(a) Turbulent kinetic energy production on the $X_{40}$ mesh. (b) Turbulent kinetic energy transport on the $X_{40}$ mesh. (c) Viscous diffusion on the $X_{40}$ mesh. (d) Dissipation on the $X_{40}$ mesh. (e) Pressure diffusion on the $X_{40}$ mesh.}
\end{figure}

The implemented methods rely on a Runge-Kutta based solver, whereas \emph{pisoFoam} relies on the temporal integration methods implemented in OpenFOAM. To make a fair comparison between these methods the Backward Euler temporal integration was chosen, since it is one of few schemes available to both solvers. Therefore, the results on the coarse mesh, shown in figures \ref{fig:CF_Coarse_Ux} through \ref{fig:CF_Coarse_pDiff}, contain a fairly large error, which is mainly caused by the Backward Euler time scheme. The emphasis of these figures, however, is to show the difference between the implemented methods and \emph{pisoFoam}, not the overall quality of the solution. To show the influence of the temporal error, the results for the $\theta_0$ method with a Runge-Kutta 3 scheme are shown in figures \ref{fig:CF_Coarse_Ux}, \ref{fig:CF_Coarse_tau} and \ref{fig:CF_Coarse_RMSs}. From these figures it can also be seen that even on a coarse mesh, the implemented solvers are able to attain fairly decent results.
From figure \ref{fig:CF_Coarse_Ux} it can be seen that there the mean stream-wise velocity for all solvers is generally much too high. \emph{pisoFoam} shows the highest mean stream-wise velocity, indicating that it is less able to develop turbulence on this mesh, which may be due to its dissipative nature. This finding is confirmed by figure \ref{fig:CF_Coarse_tau}, in which the viscous shear stress, $\rho\nu\frac{d\lr{u_x}}{dy}$, and the Reynolds stress, $\rho\lr{u_x'u_y'}$ are plotted. The highest value of the viscous stress and the lowest value of the Reynolds stress for the \emph{pisoFoam} solver confirms that this solver is least able to develop turbulence on this specific mesh.
The root-mean-square (RMS) velocities are depicted in figure \ref{fig:CF_Coarse_RMSs}, which shows that the values of $RMS(u_x)$ are too high and the RMS velocities in other directions are too low, the least accurate solution again given by \emph{pisoFoam}. These inaccuracies can be caused by coarse mesh resolution, especially in the stream-wise direction.

The turbulent kinetic energy budgets are given in figures \ref{fig:CF_Coarse_prod} through \ref{fig:CF_Coarse_pDiff}. In general, the implemented solvers show behaviour resembling the reference solution, however, because of the coarse mesh they have not attained an accurate solution. The lower turbulence of \emph{pisoFoam} can be seen in these figures by the lower absolute values of the peaks in production, transport, viscous diffusion and pressure diffusion. The dissipation budget also shows lower absolute values, and a lesser expression of the characteristic bump in the line around $y^+=15$. For all solvers, the location of the peak in production seems to occur more or less at the location where the Reynolds shear stress and the viscous stress are equal, as is expected \cite{Pope2000}. The value of the peak is slightly too high or too low for the implemented methods, with the $\theta_0$ method having the lowest production of the three. \emph{pisoFoam} has the lowest production, in line with what was seen before. The inaccuracies in production seem to be mainly compensated by the viscous diffusion around the peak and the dissipation closer to the wall. The under-estimation of the transport and viscous diffusion terms further away from the wall also seem to be compensated by the dissipation, of which the values are too high in this region. The characteristic bump in the dissipation term is not well resolved on this mesh, as the profiles are too smooth. Finally, the pressure diffusion term shows relatively large deviations from the reference value, since pressure diffusion is small in general, the deviations appear more visibly. In general, the implemented solvers show similar results, and the deviations from the reference values are expected to be caused by the coarse mesh resolution.

\begin{figure}[ht]
    \centering
    \begin{minipage}[b]{\colw\linewidth}
        \centering
        \includegraphics[height=\figh\linewidth]{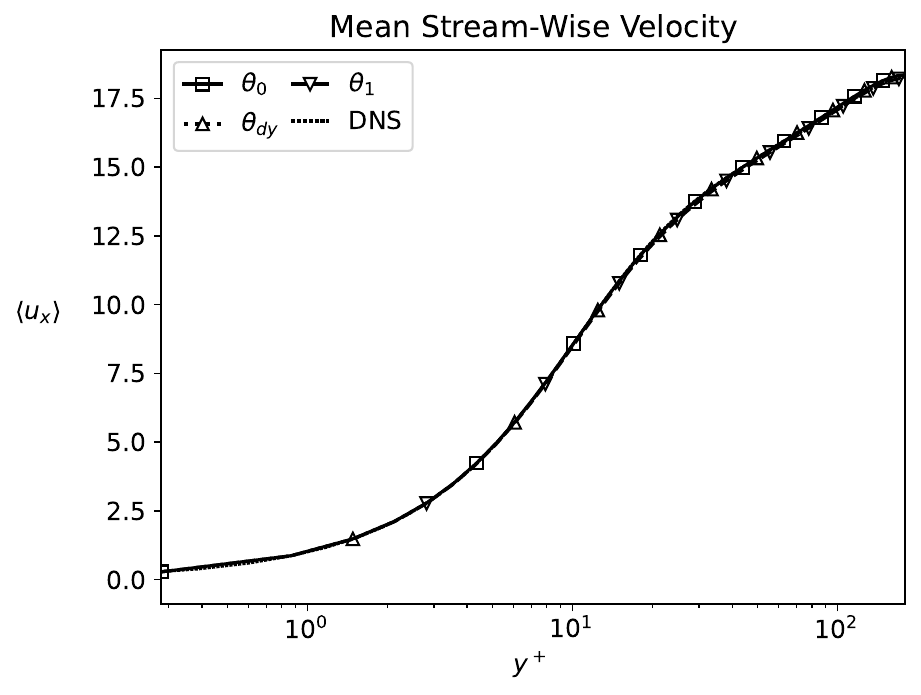}
        \subcaption{}
        \label{fig:CF_Fine_Ux}
    \end{minipage}
    \hspace{0.05\linewidth}
    \begin{minipage}[b]{\colw\linewidth}
        \centering
        \includegraphics[height=\figh\linewidth]{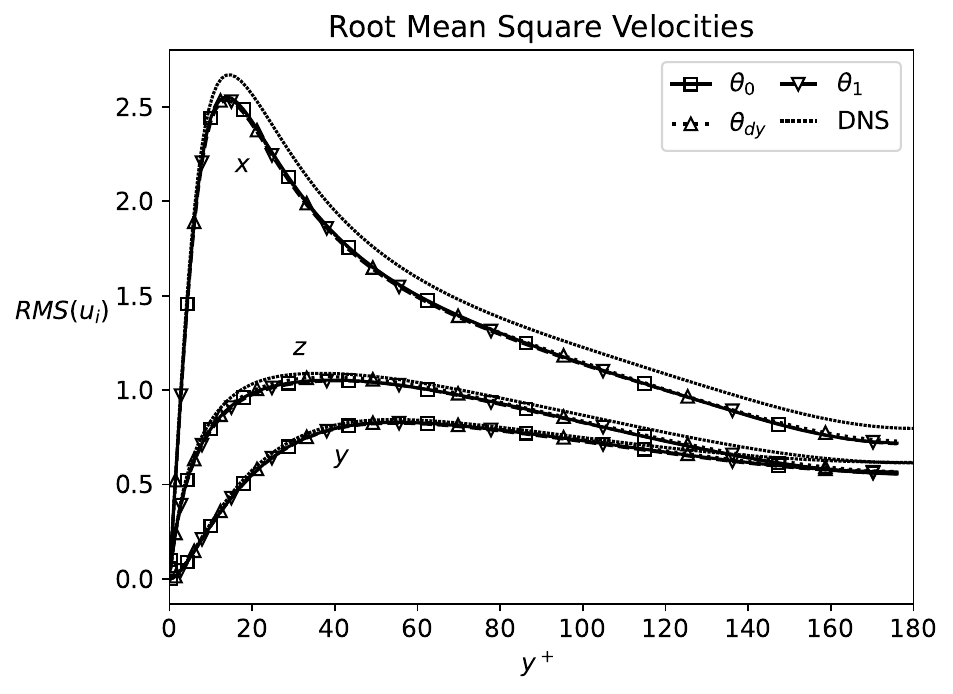}
        \subcaption{}
        \label{fig:CF_Fine_RMSs}
    \end{minipage}
    \vspace{0.05\linewidth}
    \begin{minipage}[b]{\colw\linewidth}
        \centering
        \includegraphics[height=\figh\linewidth]{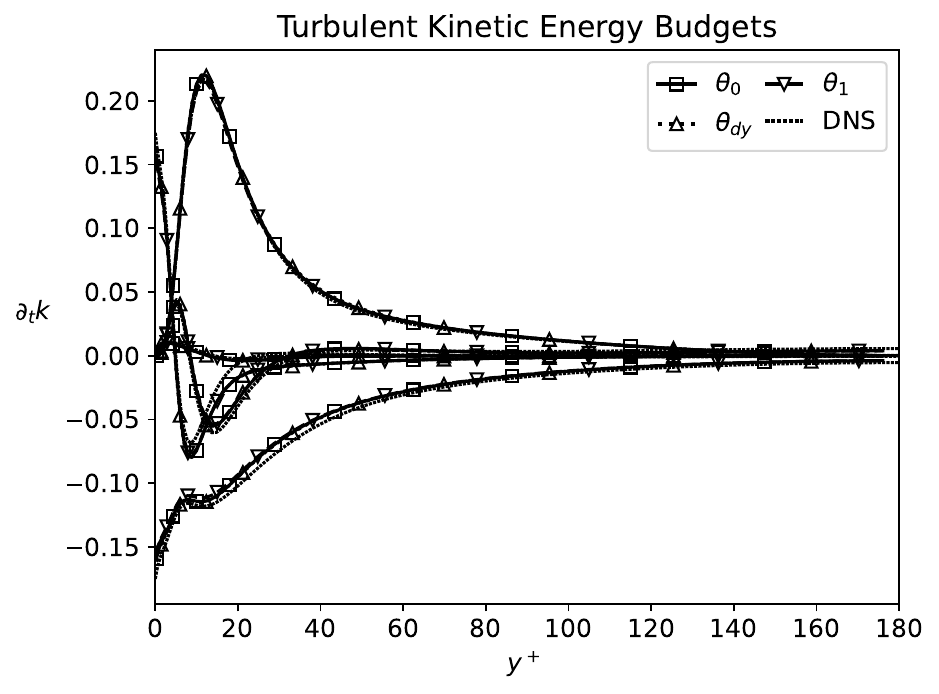}
        \subcaption{}
        \label{fig:CF_Fine_buds}
    \end{minipage}
    \caption{(a) Mean stream-wise velocity on the $X_{180}$ mesh. (b) RMS velocities on the $X_{180}$ mesh. (c) Turbulent kinetic energy budgets on the $X_{180}$ mesh.}
\end{figure}

Finally, the results for the $X_{180}$ mesh using the Runge-Kutta 3 temporal integration method are shown in figures \ref{fig:CF_Fine_Ux} through \ref{fig:CF_Fine_buds}. Figure \ref{fig:CF_Fine_Ux} shows that the mean stream-wise velocity is very accurately predicted, indicating that the turbulence is properly developed. The RMS velocities also depict this, although a slight deviation can still be seen in the values of $RMS(u_x)$. This can be explained by some under-resolution in the stream-wise direction, since the mesh is not of full direct numerical simulation (DNS) quality. Finally, figure \ref{fig:CF_Fine_buds} shows that the implemented solvers converge to one another and lie very closely to the reference values. Convergence to the reference is not yet attained for all values, but the right trend is visible and suggests that additional mesh refinement would further increase convergence of the results. Especially mesh refinement in the stream-wise direction is needed to improve the $RMS(u_x)$ values. 

Overall, with mesh refinement, the implemented solvers seem to converge to the same solution, and checkerboarding seems to diminish. The range in which the $\theta_{dy}$ method operates is therefore also smaller. On coarse meshes and in laminar flows however, the method still has significant benefits and its ability to balance dynamically between adding numerical dissipation or allowing checkerboarding. In flows that show transition from laminar flow to turbulence over time, the method should perform well with respect to the other methods. Finally, if local quantification and local balancing is developed, flows with turbulence transition areas could benefit from this method as well.

\FloatBarrier

\section{Conclusions}
\label{sec:conclusions}
This work has presented an elaborate overview of the different methods in obtaining and avoiding checkerboard oscillations when using the projection method to solve incompressible flows on collocated grids, by using the symmetry-preserving framework of which the derivations and implementations of the discrete operators were also explained in detail. 
The problems and numerical errors that accompany each method were analysed mathematically and presented in a clear manner.
The problem of the lack of definition of checkerboarding in literature was addressed and a solution to this problem was suggested by deriving a physics-based, global, non-dimensional, normalised checkerboarding coefficient. The basis of this coefficient is tied to the difference between the wide-stencil and compact-stencil Laplacian operators. This difference is linked to the fundamental problem of the collocated grid arrangement, which is unavoidably accompanied by interpolations, of which there cannot be an inverse such that the original field is reestablished, i.e. $\Gamma_{cs}\Gamma_{sc} \approx I_s$. This difference also forms the basis of the numerical errors that are introduced by the different fractional step methods to counter-act checkerboard oscillations.
This coefficient was tested using laminar and turbulent flows, and compared to qualitative and other quantitative assessments, in which it showed to give an intuitive and seemingly correct estimation of the levels of checkerboarding. 
An example of a possible usage of the coefficient was also introduced, by employing it as a parameter that gradually includes a pressure predictor into the momentum predictor equation. Since the inclusion of a pressure predictor itself can cause checkerboarding, this effectively establishes a dynamic balancing between checkerboarding and numerical dissipation through negative feedback.
In the performed test cases, this solver showed its possibility in achieving this balance without requiring any user input. By doing so it was able to achieve low levels of numerical dissipation in cases without any oscillations, whereas it was also able to reduce the amount of checkerboarding in more challenging cases. In the transient test case, the solver showed no significant difference in accuracy, while also showing convergence to the same solution as the Chorin and Van Kan methods. One drawback of this method to consider is the loss of its second-order relation between time-step size and pressure error, for which it could be interesting to consider a different feedback mechanism instead of the proposed linear feedback, such that the method stops utilising a pressure predictor only at certain levels of checkerboarding.
Another consideration of the quantification method is that the level of checkerboarding decreases greatly when the mesh is refined. Therefore, a single measurement of the level of checkerboarding without any context can be misleading or present little use. However, it still remains as a valid quantification method where no other methods are widely-known or -used.
Currently, the method has its greatest use for laminar flows, flows on coarse meshes, cases with transition between laminar and turbulent flows or more generally, simulations that are run without knowing the resulting flow a priori. For refined meshes and turbulent flows, the checkerboarding seems to diminish in general, and the operating window of the dynamic solver becomes smaller, while the accuracy of the results are similar to the Van Kan and Chorin methods, making the method less useful.
An interesting suggestion for future works would be the measurement of checkerboarding on a local or even cell level. This could provide more insight into problematic areas, suggesting mesh refinement or even feedback mechanisms that act on a local level, for example by local inclusion of a pressure predictor or hybrid usage of a compact- and wide-stencil scheme depending on the level of checkerboarding. This could then also make this method useful in cases that have laminar to turbulent flow transitional regions.

\section*{Acknowledgements}
This work is supported by the SIMEX project (PID2022-142174OB-I00) of \emph{Ministerio de Ciencia e Innovación} and the RETOtwin project (PDC2021-120970-I00) of \emph{Ministerio de Economía y Competitividad}, Spain. J.A.H. and D.S. are supported by FI AGAUR-Generalitat de Catalunya fellowships (2023 FI\_B1 00204 and 2022 FI\_B\_00173, resp.), financed  and extended by \emph{Universitat Politècnica de Catalunya} and \emph{Banc Santander}. The numerical experiments have been conducted on the Marenostrum5 supercomputer at the \emph{Barcelona Supercomputing Center} under the project IM-2024-3-0019. The authors thankfully acknowledge these institutions.


\appendix
\section{A note on collocated gradients}
\label{sec:ApGc}
\FloatBarrier

Using the definition $G_c^{\gamma} = \Gamma_{sc}^{\gamma}G = \Omega^{-1}\Gamma_{cs}^{\gamma^T}\Omega_sG$ can be quite cumbersome in numerical analyses and in terms of implementation into codes. In this section this term is rewritten to a more intuitive notation which can be easily implemented into most codes. The starting point for this derivation is a general collocated gradient, $G_c^{\gamma}$, for which the interpolation method is not yet specified. Using the definitions from \ref{tab:SPOperators}, the following derivation is made:

\begin{equation}
\label{eq:rewriteGc}
\begin{split}
    G_c^{\gamma} &= \Gamma_{sc}^{\gamma}G, \\
    &= \Omega^{-1}\Gamma_{cs}^{\gamma^T}\Omega_sG, \\
    &= -\Omega^{-1}\sbr{I_3\otimes\Pi_{cs}^{\gamma^T}}N_f^TM^T, \\
    &= \Omega^{-1}\sbr{I_3\otimes\br{T_{fo}W_o^{\gamma} + T_{fn}W_n^{\gamma}}}S_f^T\br{T_{fn}^T - T_{fo}^T}, \\
    &= \Omega^{-1}
    \begin{pmatrix*}[l]
        - & \sbr{I_3\otimes\br{T_{fo}\br{I_m-W_n^{\gamma}}}} & S_f^T & T_{fo}^T \\
         & \sbr{I_3\otimes\br{T_{fo}W_o^{\gamma}}} & S_f^T & T_{fn}^T \\
        - & \sbr{I_3\otimes\br{T_{fn}W_n^{\gamma}}} & S_f^T & T_{fo}^T \\
         & \sbr{I_3\otimes\br{T_{fn}\br{I_m-W_o^{\gamma}}}} & S_f^T & T_{fn}^T
    \end{pmatrix*}, \\
   &= \Omega^{-1}\sbr{I_3\otimes\br{T_{fo} - T_{fn}}}S_f^T\br{W_n^{\gamma}T_{fo}^T + W_o^{\gamma}T_{nf}}, \\
   &= \Omega^{-1}\sbr{I_3\otimes M}N_f^T\br{W_n^{\gamma}T_{fo}^T + W_o^{\gamma}T_{nf}}, \\
   &= G_G\overline{\Pi}_{cs}^{\gamma},
\end{split}
\end{equation}

\noindent where at each cell $i$ the following simplification is made:

\begin{equation}
    \sbr{-\sbr{I_3\otimes T_{fo}I_m}S_f^TT_{fo}^T + \sbr{I_3\otimes T_{fn}I_m}S_f^TT_{fn}^T}_i = -\sum_{f\in F(i)}\mathbf{s}_{f(i)} = \mathbf{0},
\end{equation}

\noindent which follows from the fact that the sum of all outward-pointing surface vectors of a closed geometry equals zero, which holds for any number of dimensions. In the final line, two new operators are introduced, the Gauss gradient operator, $G_G$, and the inverse-weighted cell-to-face interpolator, $\overline{\Pi}_{cs}^{\gamma}$. The Gauss gradient can be explained as assigning the normal direction of each face to the scalar value at the face, summing these face-vectors to the cell-center, then dividing by the cell volume, which is a common method to take gradients in the finite volume method. The inverse-weighted cell-to-face interpolator operates exactly as the normal cell-to-face interpolator, but the weights of neighbour and owner are swapped. This has the noteworthy property that:

\begin{equation}
    \Pi_{cs}^M = \overline{\Pi}_{cs}^M, \quad 
    \Pi_{cs}^V = \overline{\Pi}_{cs}^L, \quad
    \Pi_{cs}^L = \overline{\Pi}_{cs}^V,
\end{equation}

\noindent such that finally:

\begin{equation}
\label{eq:GcToGG}
    G_c = \Gamma_{sc}^VG = G_G\Pi_{cs}^L.
\end{equation}

\subsection{A note on implementation}
Aside from simplifying numerical analyses, equation \eqref{eq:GcToGG} greatly simplifies the implementation of the collocated gradient into code. For example, in the popular open-source finite volume code OpenFOAM, implementing $G_c$ as $\Gamma_{sc}G$ is not straight-forward, especially in symmetry-preserving codes. For these codes, the choice of interpolation has to be hard-coded into the solver, to remove any degree of freedom in choice of interpolator. Instead of writing a function for the face-to-cell interpolator, $\Gamma_{sc}$, equation \eqref{eq:GcToGG} can be used, resulting in more conventional functions that are usually readily available. For example, the symmetry-preserving OpenFOAM solver \emph{RKSymFoam} uses the following syntax to hard-code $G_c^V\mathbf{p}_c^n$ as $G_G\Pi_{cs}^L\mathbf{p}_c^n$ \cite{Hopman2023b}: 

\lstset{language=C++}
\begin{lstlisting}
const volVectorField gradpn
(
    fv::gaussGrad<scalar>::gradf
    (
        linear<scalar>(mesh).interpolate(pn),
        "gradpn"
    )
);
\end{lstlisting}

\noindent which uses the \emph{gradf()} and \emph{interpolate()} functions that are provided in the source code of OpenFOAM.

\FloatBarrier

\section{Kernel vectors for arbitrary Cartesian meshes}
\label{sec:ApKerVecGeneral}
\FloatBarrier

In this section, a set of kernel vectors for arbitrary Cartesian meshes is derived for the linear, midpoint and volumetric wide-stencil Laplacian operators, $L_c^L$, $L_c^M$, $L_c^V$. The derivation is shown for the two-dimensional case. In this case, four linearly independent kernel vectors are required to span the null-space; the constant kernel vector, $\mathbf{p}_{c(00,\gamma)}^-$, the vertical kernel vector, $\mathbf{p}_{c(10,\gamma)}^-$, the horizontal kernel vector, $\mathbf{p}_{c(01,\gamma)}^-$ and finally, the checkered kernel vector, $\mathbf{p}_{c(11,\gamma)}^-$. Since the case for $\gamma=M$ was given in \cite{larsson2010}, only the cases for $\gamma=L$ and $\gamma=V$ will be derived here. For the derivations, the fact that $Ker(G_c^{\gamma}) \in Ker(L_c^{\gamma})$ was used in addition to the equality $G_c^{\gamma} = G_G\overline{\Pi}_{cs}^{\gamma}$, which was derived in equation \eqref{eq:rewriteGc}. After applying these equalities, a set of vectors, $\pmb{\phi}_c$ needs to be found such that:

\begin{equation}
    G_c^{\gamma}\pmb{\phi}_c = G_G\overline{\Pi}_{cs}^{\gamma}\pmb{\phi}_c = \mathbf{0}_c.
\end{equation}

\begin{figure}
    \centering
    \includegraphics[width=0.8\textwidth]{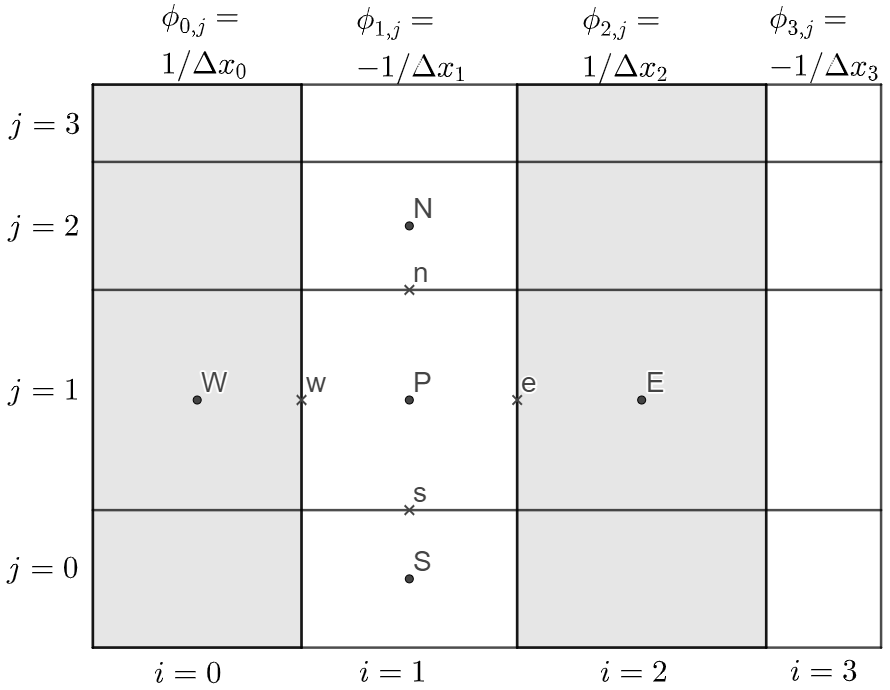}
    \caption{Vertical linear kernel vector, $\mathbf{p}_{c(10,V)}^-$, with $I=0$, $J=0$ and $\gamma=V$.}
    \label{fig:kervecVerLin}
\end{figure}

\noindent Since the constant kernel vector is a trivial solution to this equation, the kernel vector that is considered first is the vertical linear kernel vector, $\mathbf{p}_{c(10,L)}^-$. If the values of this vector are chosen as:

\begin{equation}
\label{eq:kervecVerLin}
     \sbr{\mathbf{p}_{c(10,L)}^-}_{i,j} = (-1)^i\br{\Delta x_i}^{-1},
\end{equation}

\noindent a vertically striped pattern will be obtained, as seen in figure \ref{fig:kervecVerLin}. Since the vertically neighbouring cells have equal values, the values at faces $s$ and $n$ are trivially given by:

\begin{equation}
    \sbr{\overline{\Pi}_{cs}^L\mathbf{p}_{c(10,L)}^-}_s = \sbr{\overline{\Pi}_{cs}^L\mathbf{p}_{c(10,L)}^-}_n = \frac{-1}{\Delta x_1}.
\end{equation}

\noindent Given the fact that $\overline{\Pi}_{cs}^L = \Pi_{cs}^V$, the value at face $w$ is calculated as:

\begin{equation}
\begin{split}
    \sbr{\overline{\Pi}_{cs}^L\mathbf{p}_{c(10,L)}^-}_w = \sbr{\Pi_{cs}^V\mathbf{p}_{c(10,L)}^-}_w &= \frac{\Delta x_0 \phi_W + \Delta x_1 \phi_P}{\Delta x_0 + \Delta x_1}, \\
    &= \frac{\Delta x_0 / \Delta x_0 - \Delta x_1 / \Delta x_1}{\Delta x_0 + \Delta x_1} = 0.
\end{split}
\end{equation}

\noindent Similarly, $\sbr{\overline{\Pi}_{cs}^L\mathbf{p}_{c(10,L)}^-}_e = 0$. Since this leads to opposing face pairs with equal values, it becomes immediately evident that $\sbr{G_G\overline{\Pi}_{cs}^L\mathbf{p}_{c(10,L)}^-}_P = 0$. This equality holds throughout the whole mesh and therefore the vector given by equation \eqref{eq:kervecVerLin} lies on the kernel of $L_c^L$. The horizontal linear kernel vector is derived similarly by swapping the axes. This will be shown by deriving the values for the horizontal volumetric kernel vector, $\mathbf{p}_{c(01,V)}^-$, in which, additionally, the values at each cell are replaced by their inverse. This kernel vector has its values given by:

\begin{equation}
\label{eq:kervecHorVol}
     \sbr{\mathbf{p}_{c(01,V)}^-}_{i,j} = (-1)^j\Delta y_j,
\end{equation}

\noindent so that:

\begin{align}
    \sbr{\overline{\Pi}_{cs}^V\mathbf{p}_{c(01,V)}^-}_w = \sbr{\overline{\Pi}_{cs}^V\mathbf{p}_{c(01,V)}^-}_e &= -\Delta y_1, \\
    \sbr{\overline{\Pi}_{cs}^V\mathbf{p}_{c(01,V)}^-c}_s = \sbr{\Pi_{cs}^L\mathbf{p}_{c(01,V)}^-}_s &= \frac{\Delta y_1 \phi_S + \Delta y_0 \phi_P}{\Delta y_0 + \Delta y_1}, \\
    &= \frac{\Delta y_1 \Delta y_0 - \Delta y_0 \Delta y_1}{\Delta y_0 + \Delta y_1} = 0, \nonumber \\
    \sbr{\overline{\Pi}_{cs}^V\mathbf{p}_{c(01,V)}^-}_n &= 0.
\end{align}

\noindent Which also leads to equal opposite face pair values. The checkered kernel vectors are slightly more difficult to see directly. For $L_c^V$, the values of kernel vector $\mathbf{p}_{c(11,V)}^-$ are given by:

\begin{equation}
\label{eq:kervecCheLin}
     \sbr{\mathbf{p}_{c(11,V)}^-}_{i,j} = (-1)^{i+j}\Delta x_i\Delta y_j,
\end{equation}

\noindent leading to:

\begin{align}
    \sbr{\overline{\Pi}_{cs}^V\pmb{\phi}_c}_w = \sbr{\Pi_{cs}^L\pmb{\phi}_c}_w &= \frac{\Delta x_1 \phi_W + \Delta x_0 \phi_P}{\Delta x_0 + \Delta x_1}, \\
    &= \frac{\Delta x_1 \br{\Delta x_0 \Delta y_1} - \Delta x_0 \br{\Delta x_1 \Delta y_1}}{\Delta x_0 + \Delta x_1} = 0, \nonumber \\
    \sbr{\overline{\Pi}_{cs}^V\pmb{\phi}_c}_e = \sbr{\overline{\Pi}_{cs}^V\pmb{\phi}_c}_s = \sbr{\overline{\Pi}_{cs}^V\pmb{\phi}_c}_n &= 0.
\end{align}

\noindent Which gives similar results for the checkered volumetric kernel vector. Extending these arguments for all combinations of patterns and interpolators and to the three-dimensional case, a full set of linearly independent kernel vectors for arbitrary Cartesian meshes can be derived. For the wide-stencil Laplacian operators $L_c^{\gamma}$ this set is given by:

\begin{equation}
\begin{split}
    \sbr{\mathbf{p}_{c(IJK,\gamma)}^-}_{i,j,k} &= (-1)^{iI+jJ+kK}\br{\sbr{\Delta x}_i^I\sbr{\Delta y}_j^J\sbr{\Delta z}_k^K}^{\alpha}, \quad \forall I,J,K \in \{0, 1\}, \\
    \alpha &=
    \begin{cases}
        -1 & \text{if \quad} \gamma = L \\ 
        0 & \text{if \quad} \gamma = M \\ 
        1 & \text{if \quad} \gamma = V
    \end{cases}.
\end{split}
\end{equation}

\FloatBarrier


\bibliographystyle{elsarticle-num} 
\bibliography{main.bib}

\begin{thebibliography}{10}
\expandafter\ifx\csname url\endcsname\relax
  \def\url#1{\texttt{#1}}\fi
\expandafter\ifx\csname urlprefix\endcsname\relax\def\urlprefix{URL }\fi
\expandafter\ifx\csname href\endcsname\relax
  \def\href#1#2{#2} \def\path#1{#1}\fi

\bibitem{Patankar1980}
S.~V. Patankar, Numerical Heat Transfer and Fluid Flow, Hemisphere Publishing Company, 1980.

\bibitem{Wesseling2009}
P.~Wesseling, Principles of Computational Fluid Dynamics, Vol.~29, Springer Science \& Business Media, 2009.

\bibitem{Ferziger2002}
J.~H. Ferziger, M.~Peri{\'{c}}, Computational Methods for Fluid Dynamics, Springer-Verlag, Berlin, Heidelberg, New York, 2002.

\bibitem{Versteeg2007}
H.~K. Versteeg, W.~Malalasekera, An Introduction to Computational Fluid Dynamics: The Finite Volume Method, Pearson Education, 2007.

\bibitem{Chorin1968}
A.~J. Chorin, Numerical solution of the {N}avier--{S}tokes equations, Mathematics of computation 22~(104) (1968) 745--762.

\bibitem{Temam1969}
R.~Temam, Sur l'approximation de la solution des equations de {N}avier–-{S}tokes par la methode des pas fractionnaires ii, Archive for rational mechanics and analysis 33 (1969) 377--385.

\bibitem{Patankar1983}
S.~V. Patankar, D.~B. Spalding, A calculation procedure for heat, mass and momentum transfer in three-dimensional parabolic flows, in: Numerical prediction of flow, heat transfer, turbulence and combustion, Elsevier, 1983, pp. 54--73.

\bibitem{Kim1985}
J.~Kim, P.~Moin, Application of a fractional-step method to incompressible {N}avier--{S}tokes equations, Journal of Computational Physics 59~(2) (1985) 308--323.

\bibitem{VanKan1986}
J.~{Van Kan}, A second-order accurate pressure-correction scheme for viscous incompressible flow, SIAM Journal on Scientific and Statistical Computing 7~(3) (1986) 870--891.

\bibitem{Harlow1965}
F.~H. Harlow, J.~E. Welch, Numerical calculation of time-dependent viscous incompressible flow of fluid with free surface, Physics of Fluids 8~(12) (1965) 2182--2189.

\bibitem{Perot2000}
B.~Perot, Conservation properties of unstructured staggered mesh schemes, Journal of Computational Physics 159~(1) (2000) 58--89.

\bibitem{Pascau2011}
A.~Pascau, Cell face velocity alternatives in a structured colocated grid for the unsteady {N}avier--{S}tokes equations, International Journal for Numerical Methods in Fluids 65~(7) (2011) 812--833.

\bibitem{Rhie1983}
C.~M. Rhie, W.~L. Chow, Numerical study of the turbulent flow past an airfoil with trailing edge separation, AIAA Journal 21~(11) (1983) 1525--1532.

\bibitem{Peric1988}
M.~Peri{\'{c}}, R.~Kessler, G.~Scheuerer, Comparison of finite-volume numerical methods with staggered and colocated grids, Computers and Fluids 16~(4) (1988) 389--403.

\bibitem{WenZhongShen2001}
W.~Z. Shen, J.~A. Michelsen, J.~N. S{\o}rensen, Improved {R}hie-{C}how interpolation for unsteady flow computations, AIAA Journal 39~(12) (2001) 2406--2409.

\bibitem{Choi1999}
S.~K. Choi, Note on the use of momentum interpolation method for unsteady flows, Numerical Heat Transfer; Part A: Applications 36~(5) (1999) 545--550.

\bibitem{Kawaguchi2002}
B.~Yu, Y.~Kawaguchi, W.~Q. Tao, H.~Ozoe, Checkerboard pressure predictions due to the underrelaxation factor and time step size for a nonstaggered grid with momentum interpolation method, Numerical Heat Transfer: Part B: Fundamentals 41~(1) (2002) 85--94.

\bibitem{yu2002discussion}
B.~Yu, W.-Q. Tao, J.-J. Wei, Y.~Kawaguchi, T.~Tagawa, H.~Ozoe, Discussion on momentum interpolation method for collocated grids of incompressible flow, Numerical Heat Transfer: Part B: Fundamentals 42~(2) (2002) 141--166.

\bibitem{Majumdar1988}
S.~Majumdar, Role of underrelaxation in momentum interpolation for calculation of flow with nonstaggered grids, Numerical Heat Transfer 13~(1) (1988) 125--132.

\bibitem{Miller1988}
T.~F. Miller, F.~W. Schmidt, Use of a pressure-weighted interpolation method for the solution of the incompressible {N}avier--{S}tokes equations on a nonstaggered grid system, Numerical Heat Transfer 14~(2) (1988) 213--233.

\bibitem{Yu2002}
B.~Yu, W.~Q. Tao, J.~J. Wei, Y.~Kawaguchi, T.~Tagawa, H.~Ozoe, Discussion on momentum interpolation method for collocated grids of incompressible flow, Numerical Heat Transfer: Part B: Fundamentals 42~(2) (2002) 141--166.

\bibitem{Guermond2006}
J.~L. Guermond, P.~Minev, J.~Shen, An overview of projection methods for incompressible flows, Computer Methods in Applied Mechanics and Engineering 195~(44-47) (2006) 6011--6045.

\bibitem{Zhang2014}
S.~Zhang, X.~Zhao, S.~Bayyuk, Generalized formulations for the {R}hie--{C}how interpolation, Journal of Computational Physics 258 (2014) 880--914.

\bibitem{Bartholomew2018}
P.~Bartholomew, F.~Denner, M.~H. Abdol-azis, A.~Marquis, B.~G.~M. Van~Wachem, Unified formulation of the momentum-weighted interpolation for collocated variable arrangements, Journal of Computational Physics 375 (2018) 177--208.

\bibitem{Felten2006}
F.~N. Felten, T.~S. Lund, Kinetic energy conservation issues associated with the collocated mesh scheme for incompressible flow, Journal of Computational Physics 215~(2) (2006) 465--484.

\bibitem{Ansys2023}
J.~E. Mattson, An Introduction to Ansys Fluent 2023, 1st Edition, SDC Publications, 2023.

\bibitem{STARCCM2023}
S.~D.~I. Software, Simcenter STAR-CCM+ User Guide v. 2306 (2023).

\bibitem{greenshields2023}
C.~Greenshields, \href{https://doc.cfd.direct/openfoam/user-guide-v11}{OpenFOAM v11 User Guide}, London, UK (2023).
\newline\urlprefix\url{https://doc.cfd.direct/openfoam/user-guide-v11}

\bibitem{archambeau2004code}
F.~Archambeau, N.~M{\'{e}}chitoua, M.~Sakiz, Code saturne: A finite volume code for the computation of turbulent incompressible flows-industrial applications, International Journal on Finite Volumes 1~(1) (2004).

\bibitem{Verstappen2003}
R.~W. Verstappen, A.~E. Veldman, Symmetry-preserving discretization of turbulent flow, Journal of Computational Physics 187~(1) (2003) 343--368.

\bibitem{Trias2014}
F.~X. Trias, O.~Lehmkuhl, A.~Oliva, C.~D. P{\'{e}}rez-Segarra, R.~W. Verstappen, Symmetry-preserving discretization of {N}avier--{S}tokes equations on collocated unstructured grids, Journal of Computational Physics 258 (2014) 246--267.

\bibitem{Komen2021}
E.~M. Komen, J.~A. Hopman, E.~M. Frederix, F.~X. Trias, R.~W. Verstappen, A symmetry-preserving second-order time-accurate piso-based method, Computers \& Fluids 225 (2021) 104979.

\bibitem{larsson2010}
J.~Larsson, G.~Iaccarino, et~al., A co-located incompressible {N}avier--{S}tokes solver with exact mass, momentum and kinetic energy conservation in the inviscid limit, Journal of Computational Physics 229~(12) (2010) 4425--4430.

\bibitem{Hopman2023d}
J.~A. Hopman, F.~Trias, J.~Rigola, On a conservative solution to checkerboarding: Examining the discrete laplacian kernel using mesh connectivity, in: ERCOFTAC Workshop Direct and Large Eddy Simulation, Springer, 2023, pp. 306--311.

\bibitem{Hopman2022a}
J.~A. Hopman, F.~X. {Trias Miquel}, J.~{Rigola Serrano}, Symmetry-preserving discretisation methods for magnetohydrodynamics, in: World Congress in Computational Mechanics and ECCOMAS Congress, Oslo, Norway, 2022.

\bibitem{Santos2022}
D.~Santos, F.~X. Trias, G.~Colomer, C.~D. P{\'{e}}rez-Segarra, an energy-preserving unconditionally stable fractional step method on collocated grids, in: World Congress in Computational Mechanics and ECCOMAS Congress, Oslo, Norway, 2022.

\bibitem{Santos2023}
D.~Santos, F.~X. Trias, J.~A. Hopman, C.~D. P{\'{e}}rez-Segarra, Pressure-velocity coupling on unstructured collocated grids: reconciling stability and energy-conservation, in: Proceedings of the Tenth International Symposium On Turbulence, Heat and Mass Transfer, Rome, Italy, 2023, pp. 259--262.

\bibitem{Yanenko1971}
N.~N. Yanenko, The Method of Fractional Steps, Springer-Verlag New York, 1971.

\bibitem{Chorin1967}
A.~J. Chorin, A numerical method for solving incompressible viscous flow problems, Journal of Computational Physics 2~(1) (1967) 12--26.

\bibitem{Hopman2023}
J.~A. Hopman, A.~Alsalti-Baldellou, F.~X. Trias, J.~Rigola, On a conservative solution to checkerboarding: Examining the causes of non-physical pressure modes, in: Proceedings of the 14th International ERCOFTAC Symposium on Engineering Turbulence Modelling and Measurements, Barcelona, Spain, 2023, pp. 599--603.

\bibitem{Larmaei2010}
M.~M. Larmaei, J.~Behzadi, T.~F. Mahdi, Treatment of checkerboard pressure in the collocated unstructured finite-volume scheme, Numerical Heat Transfer, Part B: Fundamentals 58~(2) (2010) 121--144.

\bibitem{Klaij2015}
C.~M. Klaij, On the stabilization of finite volume methods with co-located variables for incompressible flow, Journal of Computational Physics 297 (2015) 84--89.

\bibitem{Rauwoens2007}
P.~Rauwoens, J.~Vierendeels, B.~Merci, A solution for the odd-even decoupling problem in pressure-correction algorithms for variable density flows, Journal of Computational Physics 227~(1) (2007) 79--99.

\bibitem{Date2003}
A.~W. Date, Fluid dynamical view of pressure checkerboarding problem and smoothing pressure correction on meshes with colocated variables, International Journal of Heat and Mass Transfer 46~(25) (2003) 4885--4898.

\bibitem{Golub1996}
G.~H. Golub, C.~F. {Van Loan}, Matrix Computations, Johns Hopkins University Press, 1996.

\bibitem{Hopman2023b}
J.~A. Hopman, E.~M.~A. Frederix, \href{https://github.com/janneshopman/RKSymFoam}{{RKS}ym{F}oam {G}it{H}ub page} (2023).
\newline\urlprefix\url{https://github.com/janneshopman/RKSymFoam}

\bibitem{Vuorinen2014}
V.~Vuorinen, J.-P. Keskinen, C.~Duwig, B.~J. Boersma, On the implementation of low-dissipative {R}unge--{K}utta projection methods for time dependent flows using {O}pen{FOAM}{\textregistered}, Computers \& Fluids 93 (2014) 153--163.

\bibitem{Taylor1937}
G.~I. Taylor, A.~E. Green, Mechanism of the production of small eddies from large ones, Proceedings of the Royal Society of London. Series A - Mathematical and Physical Sciences 158~(895) (1937) 499--521.

\bibitem{Vreman2014}
A.~W. Vreman, J.~G. Kuerten, Comparison of direct numerical simulation databases of turbulent channel flow at ${Re_{\tau} = 180}$, Physics of Fluids 26~(1) (2014).

\bibitem{Komen2020}
E.~M.~J. Komen, E.~M.~A. Frederix, T.~H.~J. Coppen, V.~D'alessandro, J.~G.~M. Kuerten, Analysis of the numerical dissipation rate of different {R}unge-{K}utta and velocity interpolation methods in an unstructured collocated finite volume method in openfoam, Computer Physics Comunications (2020).

\bibitem{Zhang2015}
H.~Zhang, F.~X. Trias, A.~Gorobets, Y.~Tan, A.~Oliva, Direct numerical simulation of a fully developed turbulent square duct flow up to ${Re_{\tau}=1200}$, International Journal of Heat and Fluid Flow 54 (2015) 258--267.

\bibitem{Durbin2011}
P.~Durbin, B.~Reif, Statistical Theory and Modeling for Turbulent Flows, Wiley \& Sons, Ltd, 2011.

\bibitem{Hopman2023c}
J.~A. Hopman, \href{https://github.com/janneshopman/runTimeChannelBudgets}{run{T}ime{C}hannel{B}udgets {G}it{H}ub page} (2023).
\newline\urlprefix\url{https://github.com/janneshopman/runTimeChannelBudgets}

\bibitem{Pope2000}
S.~B. Pope, Turbulent flows, Cambridge University Press, 2000.

\end{thebibliography}

\end{document}